\documentclass[12pt]{article}
\usepackage[toc,page]{appendix}

\usepackage{hyperref}
\usepackage{natbib}
\usepackage{subcaption}
\usepackage{chngcntr}

\usepackage[utf8]{inputenc}

\usepackage{amsmath,amssymb}
\usepackage{pgfplotstable}

\newcommand{\indep}{\rotatebox[origin=c]{90}{$\models$}}
\newcommand{\proba}{\mathbb{P}}
\usepackage{fancyhdr}
\usepackage{caption}
\usepackage{graphicx}
\usepackage{color}
\usepackage{graphicx}
    \graphicspath{ {./images/} }
\usepackage{float}
\usepackage{array}
\usepackage{longtable}
\usepackage{xcolor}
\usepackage{tikz}
    \usetikzlibrary{intersections}
\usepackage{caption}
\usepackage{csvsimple}
\usepackage{appendix}

\usepackage{xcolor}

\usepackage[left=2.5cm,right=2.5cm,top=2.5cm,bottom=2.5cm]{geometry}
\usepackage{authblk} 

\title{Simulation of extreme functionals in meteoceanic data: Application to surge evolution over tidal cycles}

\author[1]{Nathan Gorse}
\author[1]{Olivier Roustant}
\author[2]{Jérémy Rohmer}
\author[2]{Déborah Idier}
\affil[1]{{\small UMR CNRS 5219, Institut de Math\'ematiques de Toulouse, INSA, Universit\'e de Toulouse, France}}
\affil[2]{{BRGM, F-45060 Orléans, France}}

\begin{document}
\fancyfoot[L]{ }
\fancyhead[R]{ }
\maketitle
\begin{abstract}
We investigate the influence of time-varying meteoceanic conditions on coastal flooding under the prism of rare events. 
Focusing on conditions observed over half tidal cycles, we observe that such data fall within the framework of functional extreme value theory, but violate standard assumptions due to temporal dependence and short-tailed behavior.a
To address this, we propose a two-stage methodology. First, we introduce an autoregressive model to eliminate temporal dependence between cycles. 
Second, considering the model residuals, we adapt existing techniques based on Pareto processes.
This allows us to build a simulator of extreme scenarios, 
by applying inverse transformations.
These simulations
depend on an initial time series, which can be randomly selected to tune the desired level of extremes.
We validate the
 simulator performance
by comparing simulated times series with observations, through several criteria, based on principal component analysis, extreme value analysis, and classification algorithms.
The approach is applied to the surge data, on the Gâvres site, located in southern Brittany, France.
\end{abstract}
\section{Introduction}

Coastal flooding frequently makes headlines due to the severe damage it causes and its long-lasting impacts. Notable examples include Storm Johanna in 2008 and Storm Xynthia in 2010, which affected vast areas and significantly raised public awareness of this risk. Coastal flooding is particularly concerning because coastal zones are densely populated and host high-value activities \citep{dawson2009integrated}. This concern is expected to grow in the coming decades, as projections estimate that between 2.1 and 2.9 billion people will live in near-coastal areas by 2100 \citep{reimann2023population}. Consequently, improving the forecasting of such events has become a critical objective for public authorities \citep{toimil2017multi}.

Numerical hydrodynamic simulators provide essential insights into the complex relationship between coastal flooding and extreme meteoceanic conditions.
In this study, we focus on extreme surge-induced flooding \citep{chaumillon2017storm} but with the added complexity that the governing conditions are time-varying and evolve over tidal cycles (3 hours around the tidal peak). In our context, the challenge lies in the fact that the inputs to the numerical models are time-dependent functions, i.e., functional inputs, and our goal is to generate extreme time series.
To this end, we adopt a functional extreme value analysis framework  which has been applied in particular to extreme windstorms, heavy spatial rainfall and temperature \citep{de2022functional,ferreira2014generalized}.

To simulate extreme data, we aim to construct a probabilistic model for these time series. However, applying the theory of functional extremes typically requires two main assumptions:
\begin{enumerate}
    \item The observations must be independent and identically distributed (i.i.d.).
    \item The time series should exhibit regular variation \citep{hult2005extremal}, implying heavy-tailed marginal distributions.
\end{enumerate}
Unfortunately, these assumptions are hardly met in many situations of interest for coastal and ocean engineering; see for instance the study by \cite{Mackay2021} for the problem of dependence, and see that of \cite{Sando2024} for an example where the extreme distributions of waves do not exhibit regular variation. This is also the case for the surge time series analysed in our case in France (full details are provided below in Sect. 2), which motivated this study.\\
To overcome this, we propose to proceed in two steps:
\begin{enumerate}
    \item Temporal dependence is addressed using an autoregressive model, which captures inter-cycle dependence while preserving intra-cycle structure.
    \item To recover heavy-tailed marginals, we apply the approach of \cite{opitz2021semi}, combining a semi-parametric model of the empirical distribution with a Fréchet marginal transformation.
\end{enumerate}

Following this preprocessing stage, the transformed time series satisfy the requirements for functional extreme modeling. We then build a probabilistic model  following the framework in \cite{dombry2015functional}, which relies on a polar coordinates representation \citep{kokoszka2019principal} whose components asymptotically follow a Pareto process. 

This probabilistic model allows us to simulate extreme time series by first sampling from that Pareto process, then applying inverse transformations to recover time series in the original space. These simulations depend on an initial time series, which is required to invert the autoregressive model. The desired extremeness level can be tuned by the choice of this reference series, enabling a range of applications -- from data-like simulations, following the same law as extreme observations, as sequences of consecutive extreme events.

A special care is paid to assess the quality of our simulated extremes by proposing a series of tools including Principal Component Analysis (PCA), extreme value analysis and two-sample classification tests \citep{lopez2016revisiting}. The overall performance of the method is evaluated using the storm surge case study.

This article is organized as follows.
In Section 2, we describe the context of our case study and analyze the characteristics of the observations.
Section 3 details the two-stage methodology leading to the probabilistic model on functional data.
Section 4 explains how this model is employed to simulate extreme data.
Section 5 applies the full framework to the storm surge case study and evaluates the simulation results.
\section{A motivating case study}
\subsection{Data description}
\begin{figure}[H]
    \centering
    \includegraphics[width=0.6\textwidth]{ 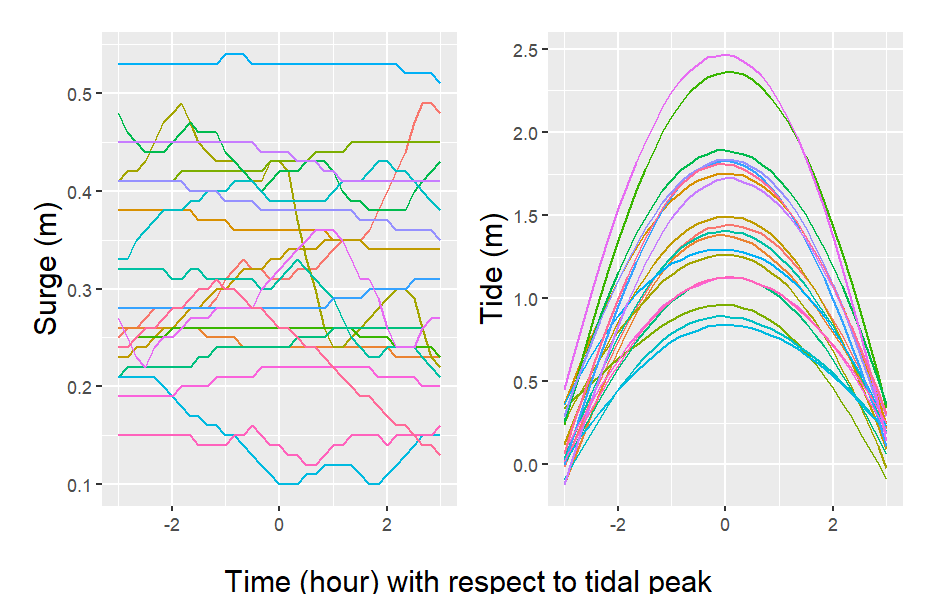}
    \caption{Sampling of $20$ observations from $1979$ to $2016$ for surge S (left) and for tide (right).}
    \label{ex_seriesTemps}
\end{figure}

We study the site of Gâvres, located on the French Atlantic coast, which is particularly sensitive to coastal flooding. The town has been impacted several times in its history like in $1904$ or in $1924$. 
Despite the evolution of its coastal defenses, about $120$ houses were flooded during storm Johanna $(08/03/2008-10/03/2008)$
\cite{origine_donnees}. Thus, it is a relevant study case to analyze extreme meteoceanic conditions. To this end, we use the hindcast database presented in \citet{origine_donnees}.\\
The small town lies in a macrotidal area such that coastal floods are strongly controlled by tides and occur around the high tide. Hence we concentrate on meteoceanic conditions over half tidal cycles, occurring within $\pm 3$ hours around high tide. Since we use a fixed time step of 10 minutes, we have several time series of length 37 for each tidal cycle. 
We consider the univariate case with an analysis of the surge (S) over the period ($1979$-$2016$) with approximately $27000$ time series that will be what we call ``observation'' in what follows.
\\
Traditionally, analyses focus on time series that follow a consistent pattern. 
For instance, in \citet{tendijck2024temporal}, all the time series considered are centered around a peak of significant wave height. 
On the contrary, the time series in our dataset do not exhibit a specific shape (Fig.~\ref{ex_seriesTemps}). 
We will use the notation $X_{M}^{t}$ to describe the value obtained at time $t$ for the $M^\text{th}$ tidal cycle. Thus,
$X_{M}^{1}, \dots, X_{M}^{37}$ represents the time series associated with the $M^\text{th}$ tidal cycle.
We study the time series that distinguish themselves by an extreme behavior. This trait refers to the framework of extreme functionals, which is based on some assumptions discussed in the next section.
\subsection{Two issues with the standard assumptions}\label{departure_std_assumptions}
\begin{figure}[H]
      \centering
      \begin{minipage}[c]{0.48\linewidth}
         \includegraphics[width=\textwidth]{ 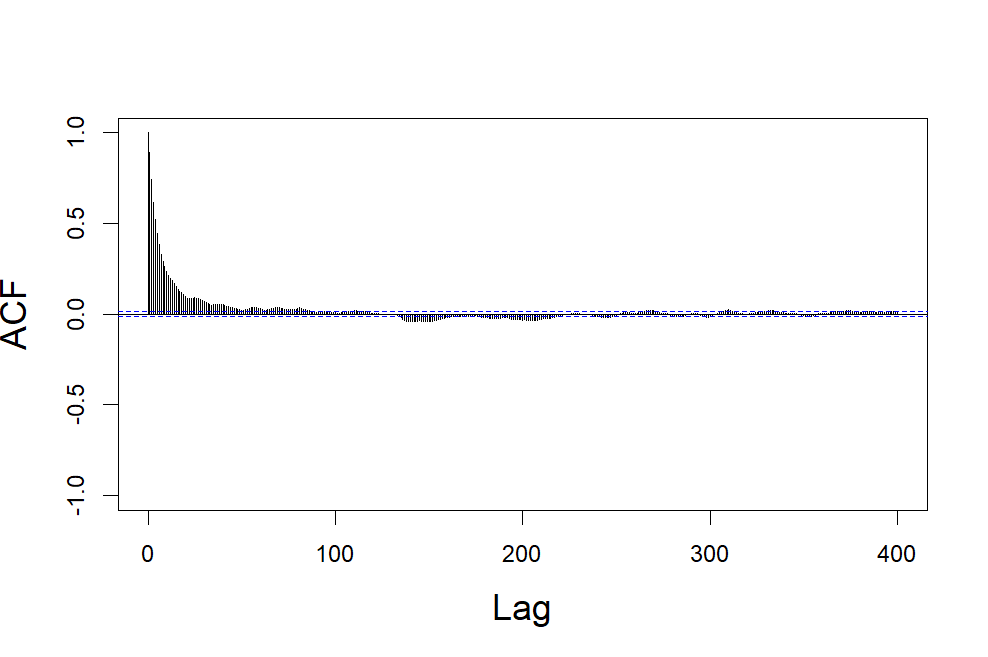}
      \end{minipage}
      \hfill
      \begin{minipage}[c]{0.48\linewidth}
         \includegraphics[width=\textwidth]{ 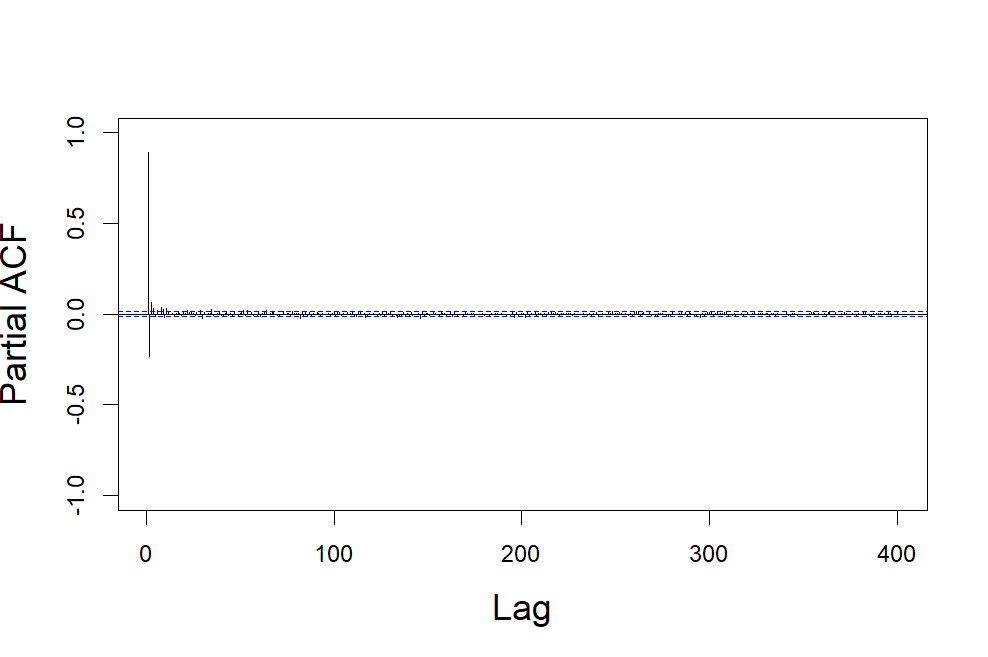}
      \end{minipage}
      \caption{Time series analysis of $S$ at the first time step of each event ($t=1$) with a lag in terms of tidal cycle index: (a) ACF; (b) PACF}
      \label{acf_pacf_S_t1}
\end{figure}

\begin{figure}[H]
    \centering
    \includegraphics[width=0.5\textwidth]{ 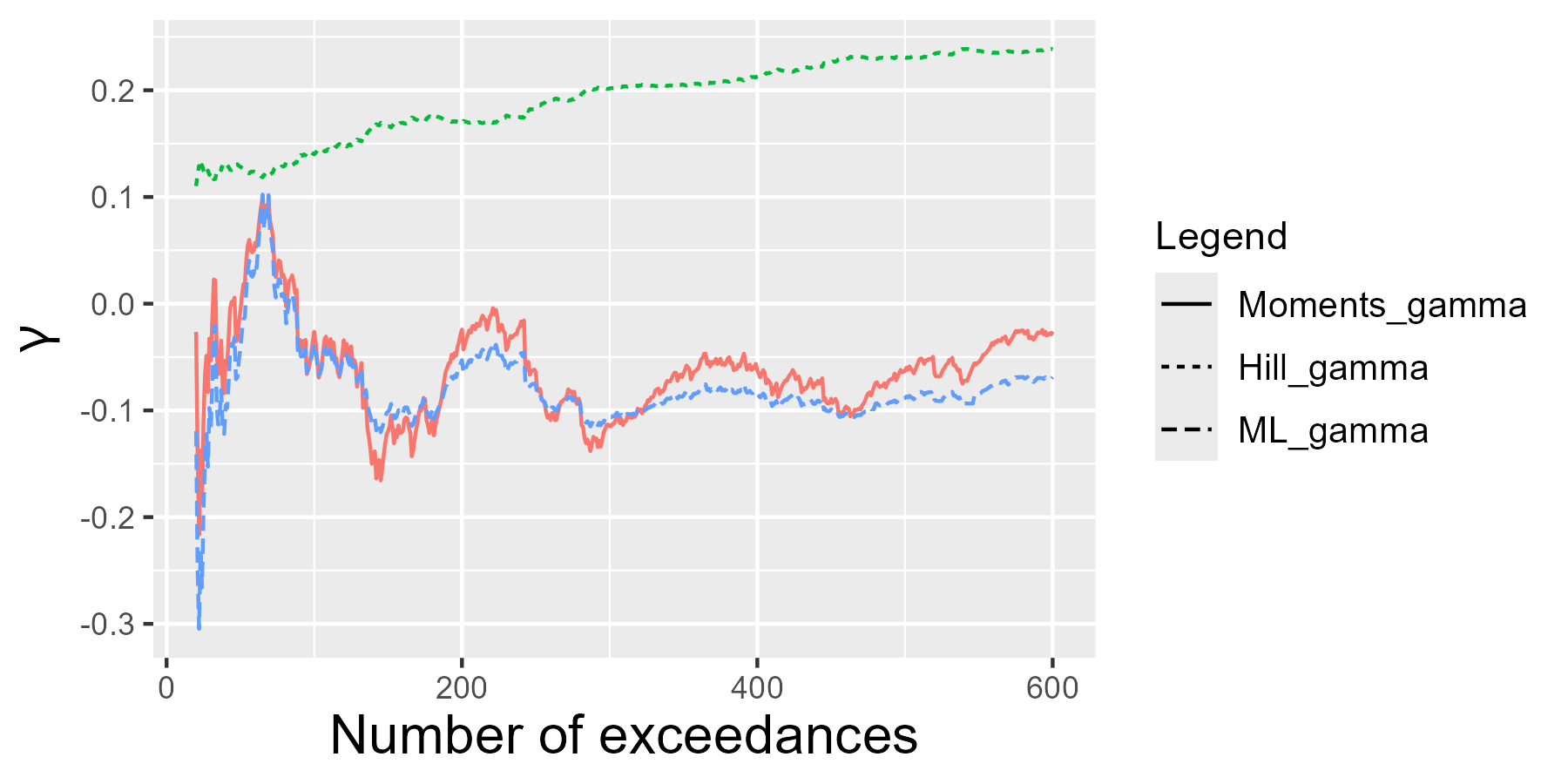}
    \caption{Estimated value of the GPD shape parameter $\gamma$ with various methods (solid line: moments estimator, dotted line: Hill estimator, dashed line: MLE estimator)}
    \label{fig:gamma_L2_raw}
\end{figure}

\paragraph{Correlated observations.}\label{corrs_plus_winter}
We generally assume that the observations (here time series of length 37) are i.i.d \citep{dombry2015functional}, which implies that, for all $t=1, \dots, 37$, the series ($X_1^{t},\dots, X_{N}^{t}$) with $N$ the number of observations should constitute an i.i.d sample. However, as illustrated by the autocorrelogram ACF and the partial autocorrelogram PACF at time $t=1$ (Fig.~\ref{acf_pacf_S_t1}), we observe that the time series are correlated for each value of $t$. The correlation level is around $0.10$ when the lag in terms of tidal cycle index equals $50$.
In addition to these strong correlations, we observe a seasonality of the variable with higher values observed during winter months (September-March) (see Appendix \ref{Seasonal}). In conclusion, the independence assumption is rejected. 
\paragraph{Departure from the regular variation hypothesis.}
The simulation method based on regular variations requires that $X_M^{t}$ is heavy-tailed distributed for each value of $t$. To properly define this concept, we recall some prerequisites in extreme value theory. Let $U$ be a univariate random variable and assume that $U$ lies in the attraction domain of a so called Generalized Extreme Value (GEV) law \citep{balkema1974residual, ferreira2014generalized}.
This means that, for a $n$-sample  $U_1, \dots, U_n$ following the law of $U$, there exists constants $a_{n}>0, b_{n}$ such that 
$$ M_{n}=\max_{i\in [[1,n]]}\frac{U_{i}-b_{n}}{a_{n}}$$ converges in distribution to a GEV law. In the Peak-Over-Threshold (POT) approach, under the same assumption, the exceedances of $U$ over a high threshold $u$ follow a Generalized Pareto distribution (GPD). Its expression is
\begin{equation}
\label{formula_GPD}
\centering
  \proba(U\leq x\mid U>u)=
  \begin{cases}
  1-\left(1+\gamma\frac{ x-u}{\sigma}\right)^{-\frac{1}{\gamma}}&  \text{if}\quad  \gamma\not =0\\
  1-\exp(-\frac{x-u}{\sigma})& \text{otherwise}
  \end{cases}
\end{equation}
where $x>u$ verifies  $1+\gamma\frac{x-u}{\sigma}>0$. We denote this law $\text{GPD}(u,\sigma,\gamma)$ and $F_{\text{GPD}}$ the corresponding cumulative distribution function (cdf), where $\gamma, \sigma$ are respectively shape and scale parameters. We say that $U$ is heavy-tailed distributed if $\gamma > 0$.\\
Following the peak-over-threshold approach \citep{coles2001introduction}, the extremeness of a time series is quantified by a homogeneous functional $\ell$, which maps the time series to a scalar value. This function is referred to as a cost functional \citep{dombry2015functional} or a risk functional \citep{de2022functional}. A time series is classified as extreme if the corresponding value of $\ell$ exceeds a specified threshold.
Using the $L^{2}$ norm as the function $\ell$, we analyze the extremes of $\ell(X_{M})$. 
\\As we estimate the evolution of $\gamma$ according to the number of exceedances, we analyze the sign of the shape parameter $\gamma$ (Fig.~\ref{fig:gamma_L2_raw}). The Hill estimator explored in \cite{de1998asymptotic} (dotted line) is based on a property of extreme observations under heavy-tail hypothesis $(\gamma>0)$. If the assumption holds, the estimator converges. 
Since the plot shows no stable region for the Hill estimator, we conclude that the condition $\gamma>0$ is not met. Hence, the regular variations assumption is not satisfied. 

\section{Construction of a probabilistic model}\label{construct_prob}
\begin{figure}[H]
    \centering
    \includegraphics[width=\textwidth]{ 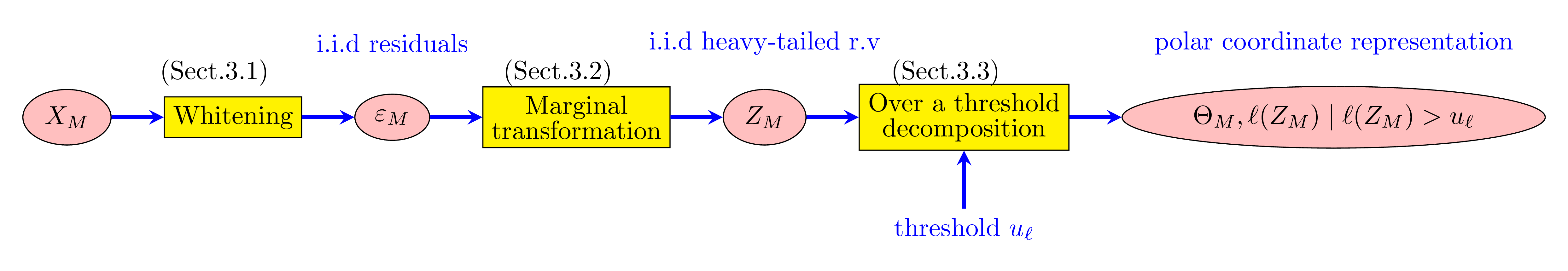}
    \caption{Construction of the polar coordinate representation from raw data (corresponding sections indicated near the boxes)}
    \label{fig:forward_pro}
\end{figure}

As discussed in the introduction, we aim at building a simulator accounting for temporal dependence and short-tailed behaviour of the observations. Our methodology has two stages. The first stage is pre-processing and aims at solving the two issues raised in the previous section. It is divided in two steps. First, we obtain independent data by considering the residuals of an autoregressive model (``whitening'' step). Second, starting from these residuals, we recover heavy-tail distributions by using marginal transformations. In a second stage, we construct a probabilistic model based on a polar coordinates representation.
The flowchart of the whole methodology is summarized in Fig.~\ref{fig:forward_pro}. 
We further detail each step in the following.
\subsection{Preprocessing stage}
\subsubsection{Recovering the independence case: Whitening with an auto-regressive model.}\label{Whiteng}
Starting from observed time series, we aim at removing the temporal dependence between successive time series. 
We first detrend the data by simply writing
\begin{align}
\label{eq:detrend}
    X_{M}^{t}=\alpha_{t}M+c_{t}+\eta_{M}^{t}=\alpha_{t}M+\tilde{X}_{M}^{t}
\end{align}
where $\alpha_{t} $ represents the slope, $c_{t}$ is the intercept and $\eta_{M}^{t}$ is the residual term. The detrended time series is denoted by $\tilde{X}_{M}^{t}$.\\
Then, we impose a minimal duration denoted $\Delta$ between each event, which implies from now on that we only consider the observations with $M=1+ \Delta m$ with $m=1,2,\dots$. 
Finally, for each value of $t$, we model the temporal dependence, with respect to $M$, by an autoregressive model, leading to
\begin{align}
\label{generalized_ar}
    \tilde{X}_{M}^{t}=\beta_{0}^{t}+\sum_{i=1}^{p}\beta_{i}^{t}
    \tilde{X}_{M-\Delta i}^{t}+\varepsilon_{M}^{t}
\end{align}
where $\beta_{0}^{t}$ is a constant term, $(\beta_{i}^{t})_{i=1,\dots,p}$ are the model parameters and $\varepsilon_{1}^{t}, \varepsilon_{1+\Delta}^{t}, \dots, \varepsilon_{N}^{t}$ are the model residuals. These residuals form a white noise, i.e. are i.i.d. random variables, that we will now consider. 
%
\subsubsection{Recovering heavy-tailed distributions: Using a Fr\'echet marginal transformation.}\label{transformation}
Starting from the residuals $\varepsilon_{M}^{t}$ obtained in the previous step, we aim at obtaining a heavy-tailed distribution.
A simple way to proceed is to use the transformation 
$\mathcal{T}:X \mapsto -1/\log(F_X(X))$ 
where $X$ is a continuous random variable with cdf $F_{X}$. Indeed, $\mathcal{T}(X)$ follows a Fréchet distribution (which is heavy tailed).
However, in our case, the cdf of $\varepsilon_{M}^{t}$ must be estimated.
Following \citet{opitz2021semi}, we approximate it by a mixture of the empirical cdf in the non-extreme part and a GPD distribution in the extreme region. Denoting $F^{\text{emp}}$ the empirical cdf and $F_{\text{GPD}}$ the distribution function of a GPD variable (\ref{formula_GPD}),  $\hat{F}_{\varepsilon^{t}}$ has the form
\begin{equation}
\label{approx_F}
  \hat{F}_{\varepsilon^{t}}(\varepsilon)=\begin{cases}
  F_{t}^{\text{emp}}(\varepsilon)&  \text{if}\quad  \varepsilon< u^{t}\\
  (1-p_{u})+ p_{u}F_{\text{GPD}(u^{t},\sigma^{t},\gamma^{t})}(\varepsilon) & \text{otherwise} 
  \end{cases}
\end{equation}
where $p_{u}$ is a parameter and $u^{t}$ verifies  $p_{u}=1-\hat{F}_{\varepsilon^{t}}(u^{t})$.
In the sequel, we will denote $Z_{M}^t=\mathcal{T}(\varepsilon_{M}^{t})$.
\subsection{Modeling stage based on a polar coordinate representation}\label{pol_rep}
Since the pre-processing stage enables us to get back to the regular variations assumptions, we can now model the law of extreme time series. 
We use the framework introduced by \citet{dombry2015functional}, in which functional data are modeled using a \emph{polar decomposition} of the form 
\begin{equation}
\label{eq:polar_decomp}
  f = \ell(f) \times A(f).  
\end{equation} where $f=Z_{M}$ in our case. The radius $\ell(f)$, chosen here  as the $L^2$ norm of $f$, allows to define extreme functionals with a scalar threshold $u_{\ell}$. Under the assumption of heavy-tailed distributions, 
the components $(\ell(f), A(f))$ become asymptotically independent, conditional on $\ell(f) > u_{\ell}$ as $u_{\ell} \to \infty$, and the conditional process $f/u_{\ell} \mid \ell(f) > u_{\ell}$ converges in distribution to a $\ell$-Pareto process.
As a consequence, we have
\begin{equation}
\label{eq:indep_seuil}
  \ell(f) \indep  A(f)| \ell(f)>u_{\ell},  \quad u_{\ell}\xrightarrow{}+\infty
\end{equation}
Thus, we must select a convenient threshold $u_{\ell}$ as defined in Eq.~\eqref{eq:indep_seuil}. One selection method is based on the convergence of the angular component $\Theta_{M}=f/\ell(f)$ of the $\ell$ Pareto process. The main idea is to analyze the behavior of $(\Theta_{M}\mid \ell(Z)>u_{\ell})$ as $u_{\ell}$ increases gradually \citep{clemenccon2024regular}. \\
After choosing a sufficiently high threshold $u_{\ell}$, we obtain the polar coordinate representation ($\Theta_{M},\ell(Z_{M})\mid \ell(Z_{M})>u_{\ell}$), which is used to simulate new extreme time series. 
\section{Simulation of extreme time series}
\begin{figure}[H]
    \centering
    \includegraphics[width=\textwidth]{ 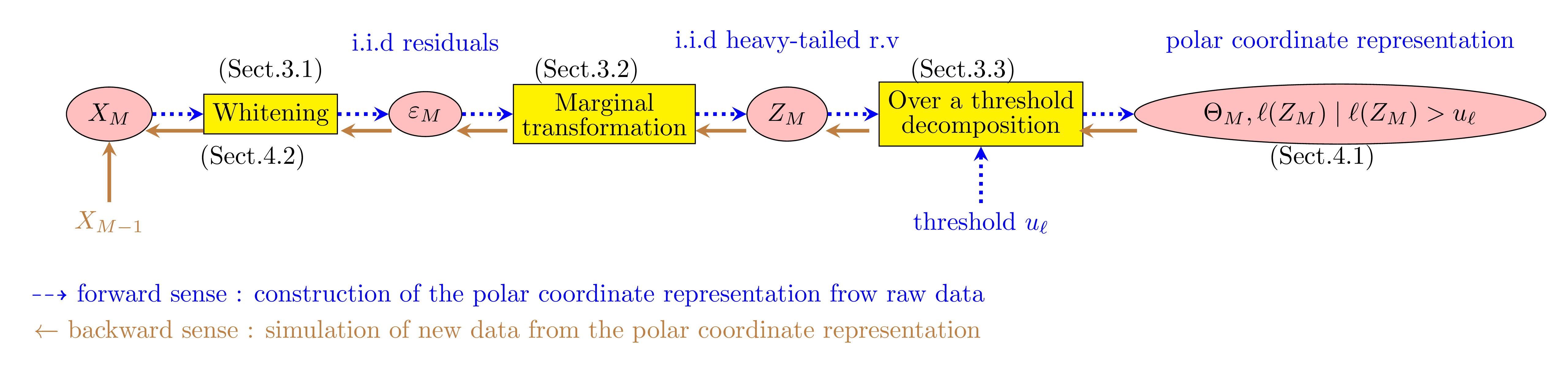}
    \caption{Two-stage methodology to simulate extreme time series (corresponding sections indicated near the boxes). The simulation starts from the left with the polar coordinate representation resulting from the forward sense (described in Sect. 3)}
    \label{fig:forward_pro_back}
\end{figure}

We propose a two-step simulator of new extreme observations based on the probabilistic model. In the first step (Sect.~\ref{ext_simul_resid}), we model the law of the polar coordinates and use properties of Pareto processes to generate simulations of $Z_{M}$, denoted $Z_{sim}$. In the second step (Sect.~\ref{simul_re}), we apply reverse transformations to $Z_{sim}$ to reconstruct time series of interest.\\
We describe the complete method in Fig.~\ref{fig:forward_pro_back}, where the forward-pointing arrows represent the transformation step introduced in Sect.~ \ref{construct_prob} whereas the backward-pointing arrows indicate the simulation phase. As previous sections describe the construction of the model, we now focus on the simulator derived from this representation. 

\subsection{Modeling of polar coordinates}\label{ext_simul_resid}
\paragraph{Functional analysis of the angular component.}\label{f_analysis_PCA}
The simulation of extreme time series is based on the polar decomposition given in Eq.~\eqref{eq:polar_decomp}. Simulating the radial component $\ell(f)$ is straightforward as its asymptotic distribution is explicit (a Pareto distribution). 
In contrast, simulating the angular component $A(f)$ is much more challenging. \\
Several simulation methods have been proposed in the literature \citep{de2018high}, which rely on a spectral representation of the coordinate-wise maximum of $Z_{M}^{t}$ using Gaussian processes \citep{dombry2015functional}. They all depend on the choice of a parametric model for the variogram. In contrast, we use a dimensionality reduction method, here PCA, before fitting parametric models. We follow \citet{clemenccon2024regular} by decomposing $\Theta_{M}$ in a orthonormal basis with
\begin{align}
\label{eq:theta_M}
    \Theta_{M}^{t}=\mu^{t}+\sum_{j=1}^{37}\langle \Theta_{M}-\mu,\nu_{j}\rangle \nu_{j}^{t}=\mu^{t}+\sum_{j=1}^{37}C_{j} \nu_{j}^{t}
\end{align}
where $(\nu_{j})_{j\geq 1}$ are eigenvectors sorted by decreasing order of the eigenvalues $(\lambda_{j})_{j\geq 1}$, $(C_{j})_{j\geq1}$ the PCA coordinates and $\mu^{t}=\mathbb{E}(\Theta_{M}^{t})$.\\
To reduce the dimensionality of the problem, we retain only the first $J$ eigenvectors and denote by $R_{J}$ the ratio of total inertia explained by the truncated basis. We truncate the PCA basis when adding another eigenvector does not significantly decrease $(1-R_{J})$.
This dimensionality reduction step allows us to estimate only the joint distribution of the scores, enabling the simulation of new extreme observations. 
\paragraph{Modeling the law of the angular component.}\label{mod_scores}
We account for the dependence between PCA coordinates by using the copula \citep{ruschendorf2009distributional} $\mathbf{C}$ defined as
\begin{align}
\label{copula_formula}
    F(C_{1},\dots, C_{J})=\mathbf{C}(F_{1}(C_1),\dots, F_{J}(C_J))=\mathbf{C}(v_1,\dots, v_J)
\end{align}
where $F$ is the joint distribution function of the vector $(C_1,\dots, C_J)$ and $F_i$ is the distribution function of the $i^{\text{th}}$ marginal. \\
We approach the joint law of $(v_{1},\dots,v_{J})$ by fitting a parametric copula model. 
As copula models are defined for bivariate settings, we use vine copulas \citep{czado2022vine} when $J>2$, which use a tree representation of the dependence structure where each edge corresponds to a bivariate copula. Then, the law of the whole vector is modeled by combining the pairwise models in a hierarchical manner. To simulate new scores $(C_{\text{sim},1},\dots, C_{\text{sim},J})$, we apply $F_i^{-1}$ for $i=1,\dots, J$ to new values $(\Tilde{v}_1,\dots,\tilde{v}_{J})$ sampled from the model.\\ 
With this method, we generate $n_{sim}$ new angles $\Theta_{\text{sim}}$ by rewriting Eq.~\eqref{eq:theta_M}, which gives:
\begin{align}
\label{simul_theta}
  \Theta_{\text{sim}}^{t}=\mu^{t}+\sum_{j=1}^{J}C_{\text{sim},j} \nu_{j}^{t}
\end{align}
As $\ell(\Theta_M)$ must equal $1$, we return $\Theta_{\text{sim}}=\Theta_{\text{sim}}/\ell(\Theta_{\text{sim}})$. This simulation step makes it possible to simulate new extreme time series, which is discussed in the next section.
\subsection{Applying the reverse transformations}\label{simul_re}
\paragraph{Over a threshold decomposition and marginal transformation.}
Thanks to the independence established in Eq.~\eqref{eq:indep_seuil}, we generate new extreme time series $Z_{sim}$ by simulating new values of the radius $\ell(f)$ and by combining them with the simulations $\Theta_{\text{sim}}$.
As $Z_{M}^{t}\sim \text{Frechet}(1)$, we impose that $Z_{sim}^{t}>0$ for every value of $t$. To this aim, we apply a rejection sampling: if a curve is strictly above $0$, we accept it; otherwise, we discard it and we simulate both a new $\Theta_{\text{sim}}$ and a new radius.  \\
The final step consists in applying the reversed transformation $\mathcal{T}^{-1}$ to the accepted time series, yielding extreme time series $\varepsilon_{\text{sim}}$, which can be used to simulate new realizations of the time series $\tilde{X}_{M}$. 
\paragraph{Inverting the autoregressive model.}\label{inversion_AR}
In the case where $p=1$, we obtain simulated time series for $(\tilde{X}_{\text{sim}}^t)_{t=1, \dots, 37}$, depending on initial time series $(\tilde{X}_{M-\Delta}^t)_{t=1, \dots, 37}$ with
\begin{equation}
\label{eq:AR_reg}
  \tilde{X}_{\text{sim}}^{t}=\beta_{0}^{t}+\beta_{1}^{t}\tilde{X}_{M-\Delta}^{t}+\varepsilon_\text{sim}^{t}
\end{equation}
The values and the shape of $\tilde{X}_{\text{sim}}$ depend on the levels reached by $\tilde{X}_{M-\Delta}$. In this context, we can use an extreme $\Tilde{X}_{M-\Delta}$ to produce consecutive extremes. As our objective is to simulate data-like extreme time series, we must choose an appropriate time series $\tilde{X}_{M-\Delta}$ given $\varepsilon_{sim}$. \\
Thus, we can adopt a conditional or an unconditional approach to sample $X_{M-\Delta}$. If we choose a conditional sampling, we account for the joint distribution of the couple $(X_{M-\Delta}^{t},\varepsilon_{M}^{t})$ as these variables may still exhibit dependence, which should be preserved during the sampling process. 
A simple approach to incorporate this dependence is to sample from the empirical distribution of $(\ell(\Tilde{X}_{M-\Delta})\mid \ell(\varepsilon_{M})=x)$. For each value $x$ of $\ell(\varepsilon_{\text{sim}})$, we define a symmetric window centered at $x$ that include $20$ neighboring data points. Then, we draw randomly with replacement one point from this window.\\
Finally, following Eq.~\eqref{eq:detrend}, we incorporate the trend of $X_{M}^{t}$ over the period by using the equation:
\begin{align}
\label{reconstuct_theta}
    X_{\text{sim},M}^{t}=\alpha_{t}M+\tilde{X}_{\text{sim}}^{t}
\end{align}
where $M$ is known thanks to the sampling of the $\tilde{X}_{M-\Delta}$. 
Another approach would be to predict today's forcing conditions by using the current index. Yet, we focus on the detrended time series to facilitate direct comparisons between our simulations and the observed extreme time series. 
We apply our method in the next section to the case study and discuss the validation of the simulator. 
\subsection{Implementation.}
The method is coded in \citep{R_core} and we used several packages. 
POT \citep{pot_package} and extRemes \citep{extremes_package} enabled us to study exceedances at each time. We based the marginal transformation code on \citep{opitz2021semi}.
Besides, we used FactoMineR \citep{Facto_package} for PCA and VineCopula \citep{Vinecopula} to approach the law of copulas. We also use the package \citep{MVT_package}. Finally, we used e1071 \citep{package_svm_RF} and randomForest \citep{RFML} to construct support-vector machines and random forest classifiers.
\section{Application to the case study}
In this section, we apply the methodology detailed in Sect.~3 and Sect.~4 to the available database from the site of Gâvres. Using the reading directions of Fig. \ref{fig:forward_pro_back}, we present some implementation details and results for the key steps of the procedure.
\subsection{Forward sense: Construction of the polar coordinate representation}
\begin{figure}[H]
    \centering
    \includegraphics[width=0.5\textwidth]{ 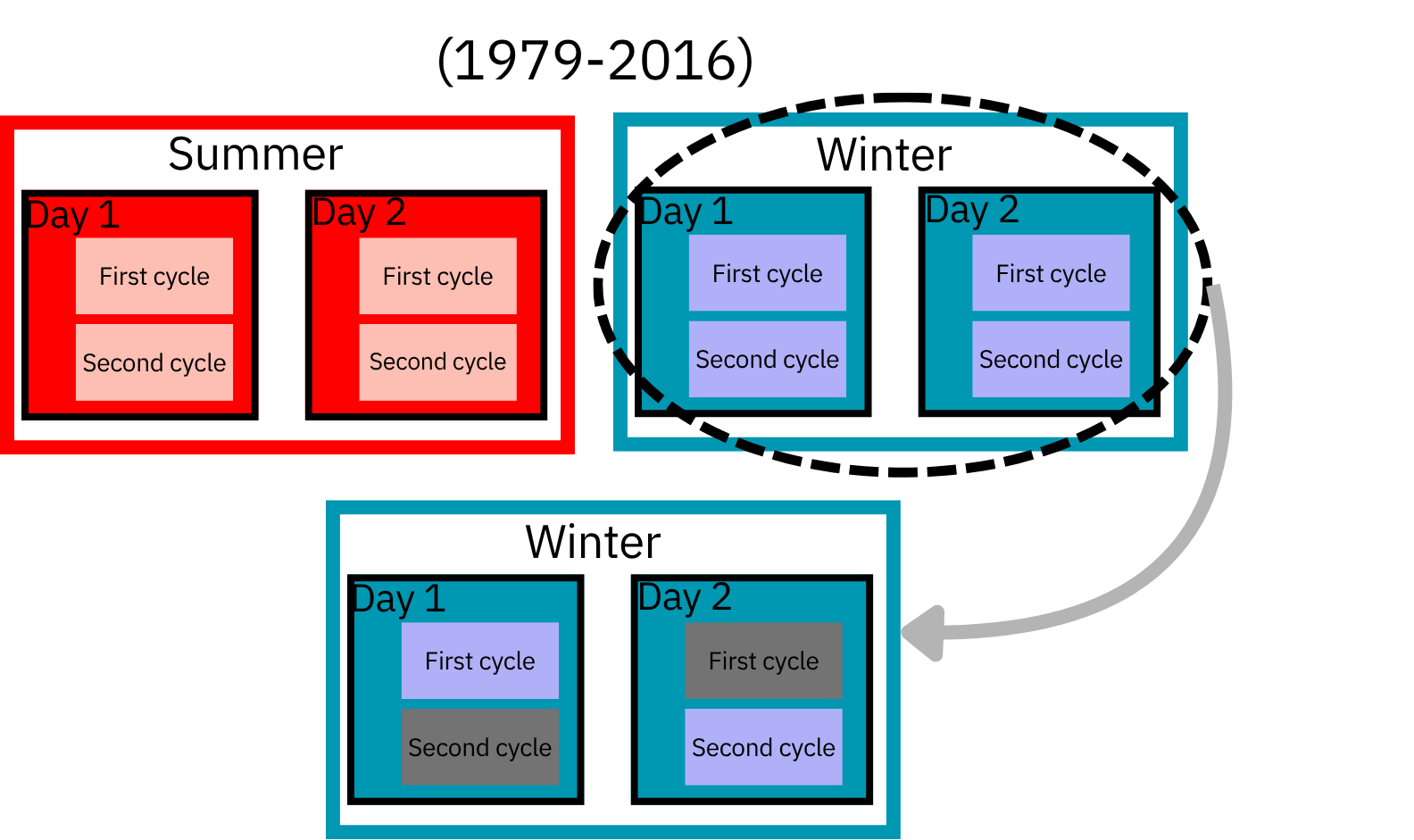}
    \caption{Summary of the database construction: selection of winter time series and one cycle out of $3$}
    \label{fig:sub_data_cstruct}
\end{figure}

\begin{figure}[H]
    \centering
    \includegraphics[width=0.7\textwidth]{ 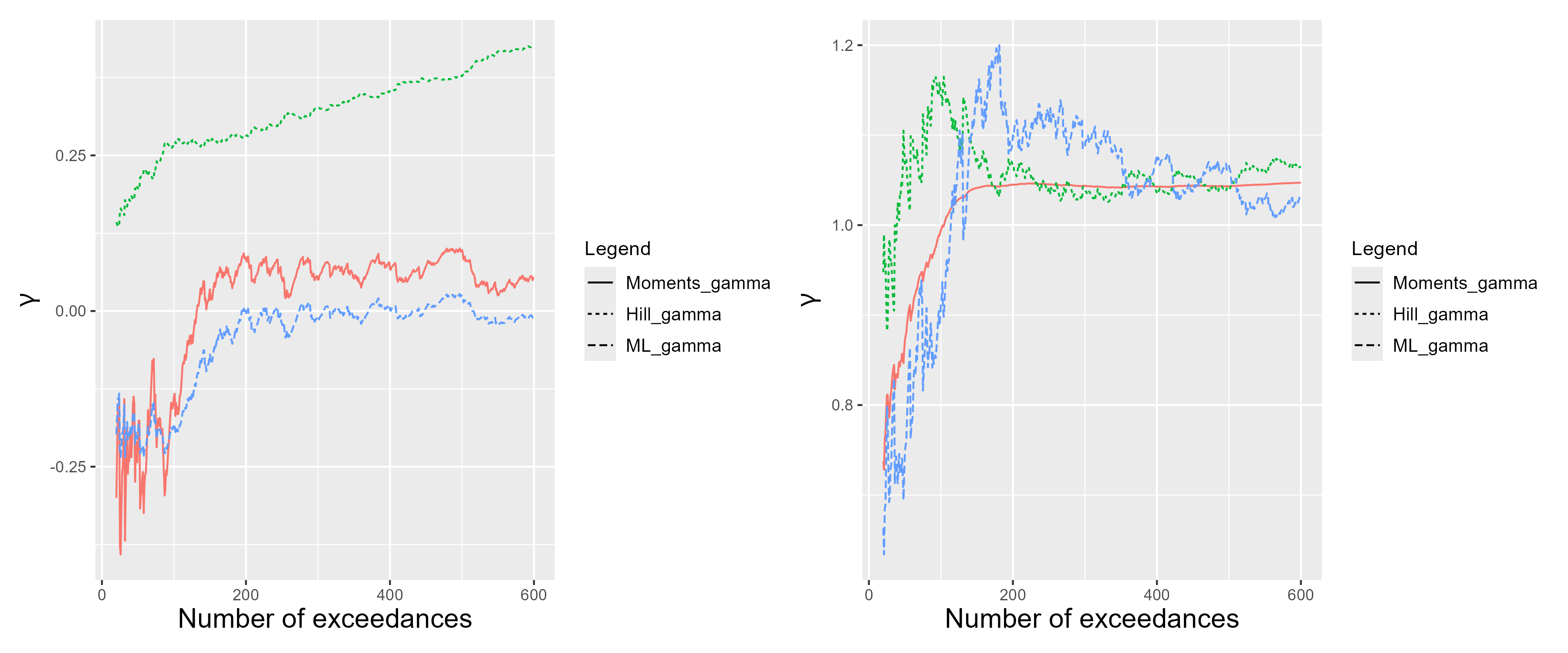}
    \caption{Estimation of $\gamma$ for: (a) $\ell(\varepsilon_{M})$; (b) $\ell(\mathcal{T}(\varepsilon_{M}))$ (solid line: moments estimator, dotted line: Hill estimator, dashed line: MLE estimator)}
    \label{effet_transform_Surcote}
\end{figure}

\begin{figure}[H]
    \centering
    \includegraphics[width=0.6\textwidth]{ 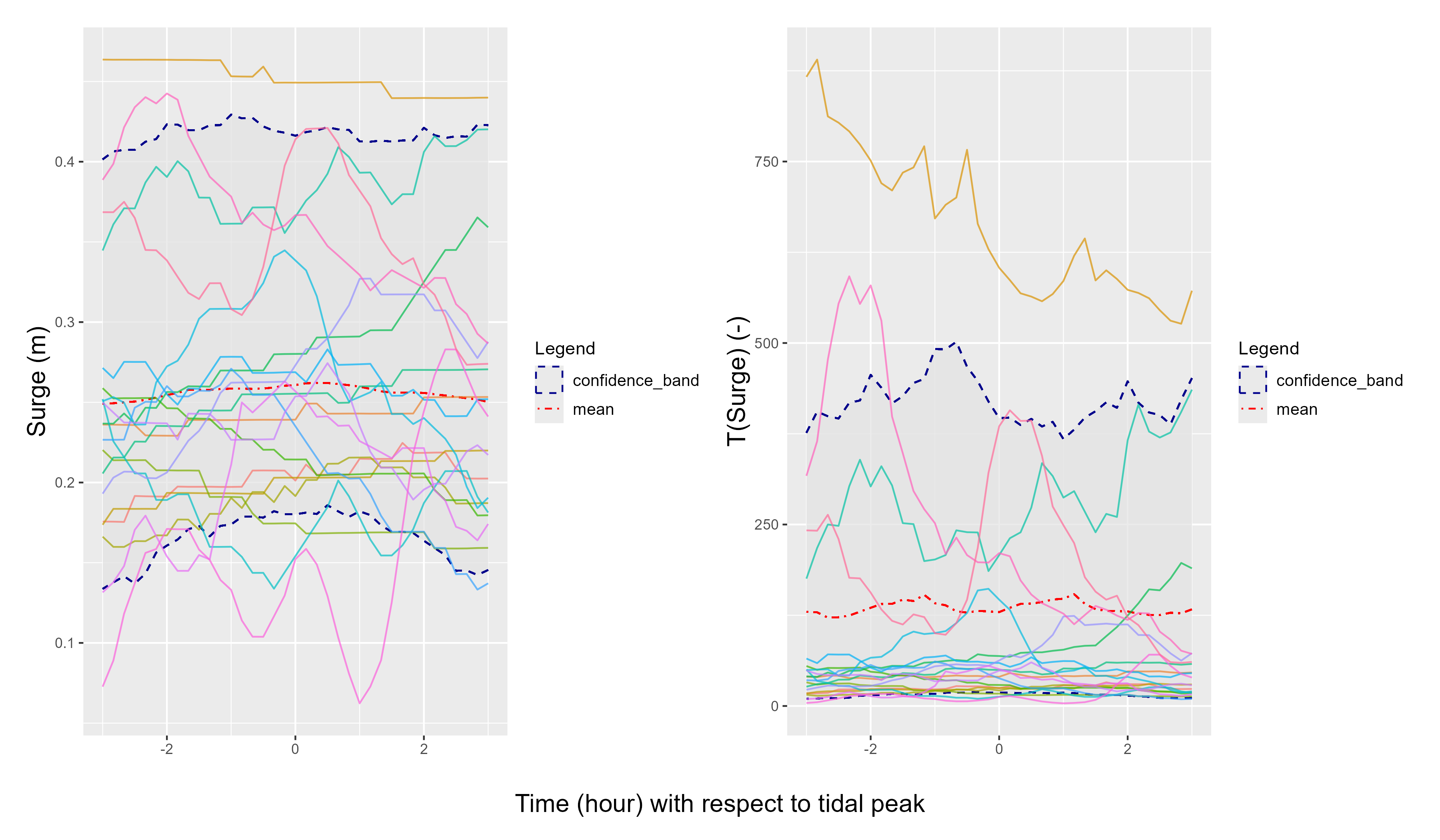}
    \caption{Sample of $20$ time series selected for the surge (left panel: original scale, right panel: Frechet scale with $95\%$ confidence intervals in dotted lines)}
\label{fig:exts_S}
\end{figure}

Following Fig.~\ref{fig:forward_pro}, we apply a two-step method to obtain the representation used for generating new extreme time series. Since our data do not respect the framework assumptions (Sect.~\ref{departure_std_assumptions}), the first step consists in retrieving the framework setting whereas the second step enables to produce the polar representation. 
\paragraph{Pre-processing stage: Whitening and marginal transformation.}\label{marg_transf_GPD}
First, by applying the whitening stage described in Sect.~\ref{Whiteng}, we detrend the time series $X_{M}^{t}$ with Eq.~\eqref{eq:detrend} and consider the detrended time series $\tilde{X}_{M}^{t}$. Since significant correlations remain, we retain only one observation out of $3$, corresponding to $\Delta=3$. The following graphic (Fig.~\ref{fig:sub_data_cstruct}) illustrates the database construction, where the grey rectangles represent the tidal cycles that are excluded from the analysis.
We now analyze the pair of correlated variables $(\Tilde{X}_{M}^{t},\Tilde{X}_{M+\Delta h}^{t})$. 
The slow decay of correlation levels on Fig.~\ref{acf_pacf_S_t1}a and the significant peak in Fig.~\ref{acf_pacf_S_t1}b at $h=1$ suggest the use of the models Eq.~\eqref{generalized_ar} with order $p=1$. Thus, we apply for each value of $t$ an autoregressive model of order $1$ denoted AR($1$), which reduces effectively the correlation value. We describe with more details the effect of these models in Appendix~\ref{Whitening_}. 
\\
Besides, since we study $\varepsilon
_{M}^{t}$'s exceedances for a given $t$, we assess extremal dependence as defined in \citet{coles1999dependence}. To this end, we apply the statistical test proposed by \cite{thomas2001statistical} whose null hypothesis is the asymptotic dependence of the residuals $\varepsilon_{M}^{t}$. The null hypothesis is rejected at confidence level $1\%$ ($p$ value $<$ $e-10$), indicating that the residuals $\varepsilon_{M}^{t}$ are asymptotically independent.\\
Then, as detailed in Sect.~\ref{transformation}, we aim at obtaining heavy-tailed marginals for each value of $t$. To achieve this, following \cite{opitz2021semi}, we estimate the cdf of $\varepsilon_{M}^{t}$ and apply the transformation $\mathcal{T}$ to the residuals, obtaining Frechet marginals $Z_{M}^t$ for each value of $t$.  \\ 
An intermediate step involves selecting the threshold parameter $u^{t}$ used in Eq.~\eqref{approx_F}. To guide this choice, we use the property that for any $(w^{t}>u^{t})$ the exceedances $(\varepsilon_{M}^{t}>w^{t})$ should follow a GPD law if $u^{t}$ is a suitable threshold. Accordingly, we define updated parameters $(\sigma',\gamma')$ that remain stable as $w^t$ increases.
With the same objective, we analyze, for each value of $t$, the behaviour of the mean residual life (MRL) $\mathbb{E}(\varepsilon^{t}_{M}-u^{t}\mid \varepsilon^{t}_{M}>u^{t})$ \citep{coles2001introduction} as a function of $u^{t}$. 
Guided by these tools, we define $u^{t}$ as the $90\%$ percentile of $\varepsilon_{M}^{t}$ since the MRL plot becomes linear beyond this point and the couple $(\sigma',\gamma')$ remains stable for all thresholds $(w^{t}>u^{t})$ (see Appendix~\ref{GPD_choice}).
\\
To assess the quality of the marginal transformation, we compare empirical quantiles $Z_{M}^{t}$ with those of the Frechet distribution across various values of $t$. The levels obtained are close to the theoretical ones, demonstrating the consistency of the transformation.\\ 
The transformation induces a change in the tail behavior of the distribution of $\ell(.)$: while $\ell(\varepsilon_{M})$ is light-tailed distributed (see Fig.~\ref{effet_transform_Surcote}a), whereas $\ell(Z_{M})$ displays heavy-tailed behavior (Fig.~\ref{effet_transform_Surcote}b), as confirmed by the convergence of the Hill estimator.
\paragraph{Modeling stage based on a polar coordinate representation.}
As we get back to the framework assumptions with the pre-processing step, we now choose the threshold $u_{\ell}$ that defines the set of extreme observations via the condition $\ell(Z_{M})>u_{\ell}$ 
of Eq.~\eqref{eq:indep_seuil}. Following Sect.~\ref{pol_rep}, the choice of $u_{\ell}$ depends on the convergence behavior of $\Theta_{M}$ when $u_{\ell}$ goes to infinity. In our case, $\Theta_{M}$ appears to converge in distribution when the number of extreme observations is between $200$ and $300$, corresponding to approximately $4\%$ and $6\%$ of the database (see Appendix \ref{theta_conv_ul}). 
\\As a consequence, we select the top $5\%$ of the dataset, corresponding to approximately $7$ extreme observations per year, and represent a sample of the resulting $n=259$ time series on their original scale and on the Frechet scale (Fig.~\ref{fig:exts_S}). The figure shows that applying the marginal transformation amplifies the variations between successive measures. 
\subsection{Backward sense: Simulation of extreme time series}
\begin{figure}[H]
      \centering
    \includegraphics[width=0.7\textwidth]{ 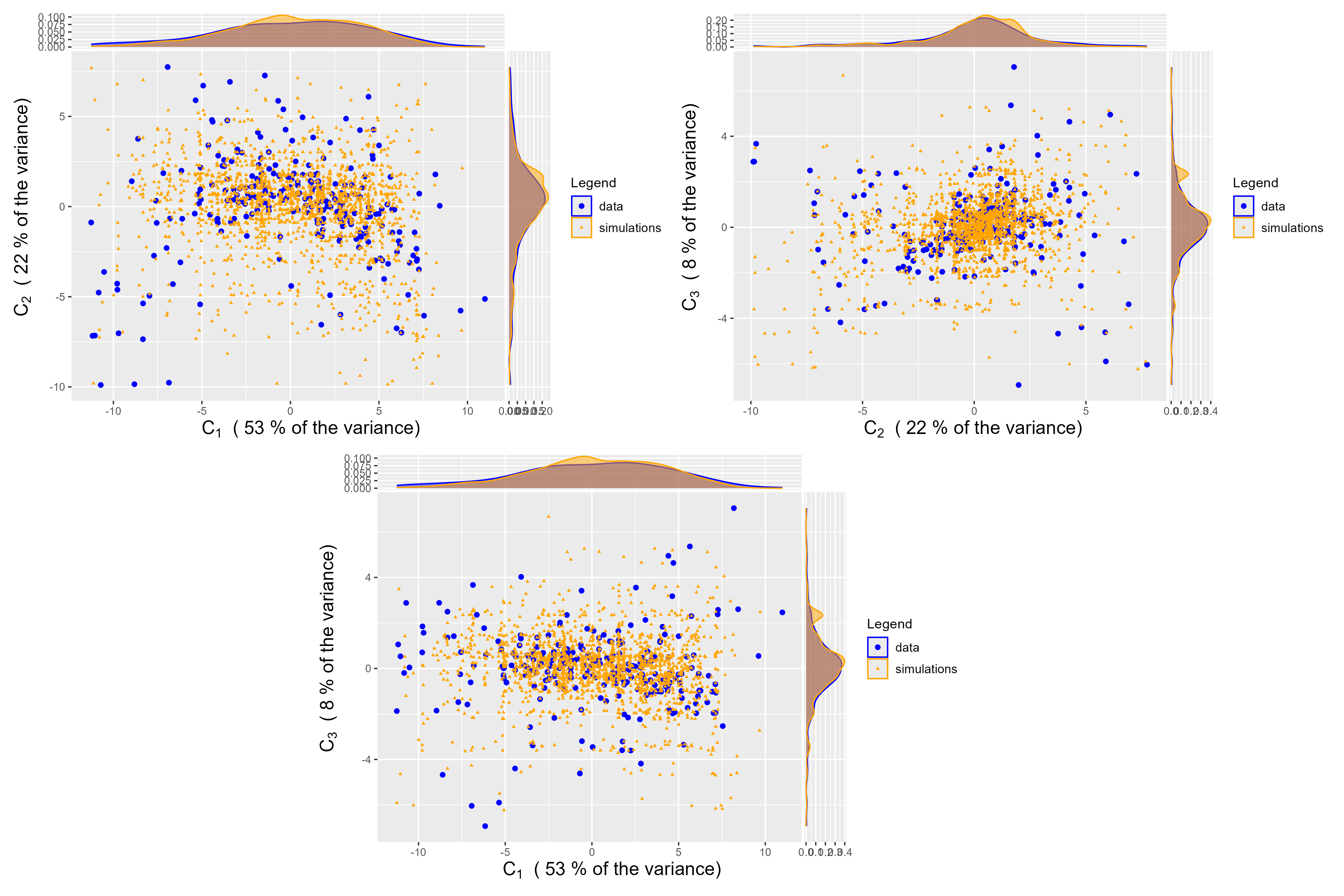}
      \caption{Coordinates obtained of $2,000$ simulated and recorded extreme time series: (a) ($C_{1},C_{2})$ ; (b) ($C_{2},C_{3}$) ; (c) ($C_{1},C_{3}$)}
      \label{fig:acp_theta_}
\end{figure}

\begin{figure}[H]
    \centering
    \includegraphics[width=0.6\textwidth]{ 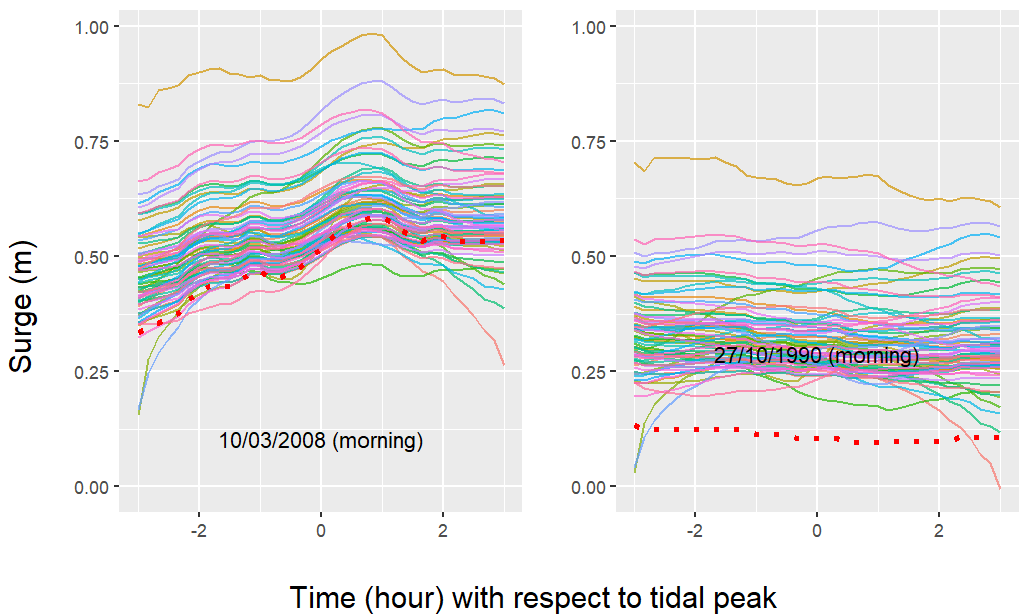}
    \caption{Simulated time series obtained with two different types of $\tilde{X}_{M-\Delta}$ (dotted line) selected among extreme observations (left panel) or non-extreme ones (right panel).}
    \label{init_ext_or_not}
\end{figure}

The transformations of the previous section enable us to use a convenient probabilistic model for extreme time series. Following Fig. \ref{fig:forward_pro_back}, we now describe the simulator based on this model, which uses the reverse transformations. 
\paragraph{Modeling the law of polar coordinates.}
We use the polar decomposition defined in Eq.~\eqref{eq:polar_decomp} as a basis for our simulation method. After selecting the $n=259$ extreme observations, we first compute their angular component $\Theta_{M}$ and derive the corresponding PCA basis Eq.~\eqref{eq:theta_M}. Following Sect.~\ref{f_analysis_PCA}, we select the first $J=3$ eigenvectors of the basis as the ratio $(1-R_{J})$ slowly decreases  beyond the third dimension (see Appendix C: Fig.~\ref{fig:prop_inertia}). With this choice, the selected dimensions explains $R_J=83\%$ of the variance. \\
We use vine copulas and model the joint distribution of the coordinate vector by fitting a range of copula families, namely elliptical and Archimedean copulas such as Tawn and Clayton models, to the bivariate vectors defined by the vine structure.
Finally, for each pair of coordinates, we use the Akaike Information Criterion (AIC) \citep{akaike1974}
to select the best copula and estimate the tail coefficients $(\lambda_{+},\lambda_{-})$ along with the Kendall's $\tau$. In our case, a $t$ copula is selected for each pair of coordinates. Please refer to Appendix~\ref{details_cop} for additional information about the modeling. \\
The iso-density contours of the fitted law are consistent with the distribution of the data points. To go further, as we assume that the observations $(v_{1},\dots, v_{J})$ follow the fitted copula model, we perform some goodness-of-fit tests presented in \citet{schepsmeier2019goodness}. 
We do not reject the null hypothesis when we use the bootstrap test based on the White statistic \citep{white1982maximum} ($p$ \text{value} $\approx 0.5$). \\
Using the composite copula model, we simulate new vectors $(v_{sim,1},v_{sim,2},v_{sim,3})$ and transform each coordinate back to its initial scale by applying $F_{i}^{-1}$ for $i=1,\dots, 3$. We simulate $n_{sim}=2,000$ triplets of coordinates from the model and we compare the simulated values with the coordinates of the extreme observations (Fig.~\ref{fig:acp_theta_}). 
The marginals obtained in the simulations are consistent with the data distributions, indicating that our model is consistent with the dependence structure of the coordinate vector.
\\
We use the coordinates $C_{\text{sim}}$ to obtain the angular components $\Theta_{\text{sim}}$ by applying Eq.~\eqref{reconstuct_theta}. The resulting patterns resembles that of the extreme observations although the simulated cuves appear to be smoother.
\paragraph{Simulating new extreme times series: Applying reverse transformations}
Following Sect.~\ref{simul_re}, we generate extreme time series $Z_{sim}$ by combining the simulated angles $\Theta_{\text{sim}}$ with simulations of $\ell(Z)$. Then, we apply the inverse transformation $\mathcal{T}^{-1}$ to $Z_{sim}$ to return to the original scale, which produces simulated extreme residuals $\varepsilon_{sim}$. \\ 
Using Sect.~\ref{inversion_AR}, we now invert the estimated AR($1$) models to return simulated time series $\tilde{X}_{sim}$. We consider two choices of $\tilde{X}_{M-\Delta}$: one observed during the extreme storm Johanna ($10/03/2008$) and another with a non-extreme behavior ($27/10/1990$), verifying $\ell(\mathcal{T}(\varepsilon_{M}))<u_{\ell}$.
The results, displayed in Fig.~\ref{init_ext_or_not}, show that the choice of $\tilde{X}_{M-\Delta}$ significantly affects both the values and the shape of $\tilde{X}_{\text{sim}}$. The left panel reveals notably high values, illustrating the effect of having consecutive extremes. 
Since the variables ($\tilde{X}_{M-\Delta}^{t},\varepsilon_{M}^{t}$) are dependent, we apply the conditional sampling illustrated in Sect.~\ref{inversion_AR} (see Appendix~\ref{choice_previous_obs}). Then, we generate new time series $\tilde{X}_{\text{sim}}$ by using $\varepsilon_{\text{sim}}$ and the selected $\tilde{X}_{M-\Delta}$. We see that the shape and the values of simulated and observed extreme time series are quite similar, suggesting that our simulations are consistent with the observations.  
However, we can use other methods to validate our simulator, which is discussed in details in the following. 
\begin{figure}[H]
    \centering
    \includegraphics[width=0.6\textwidth]{ 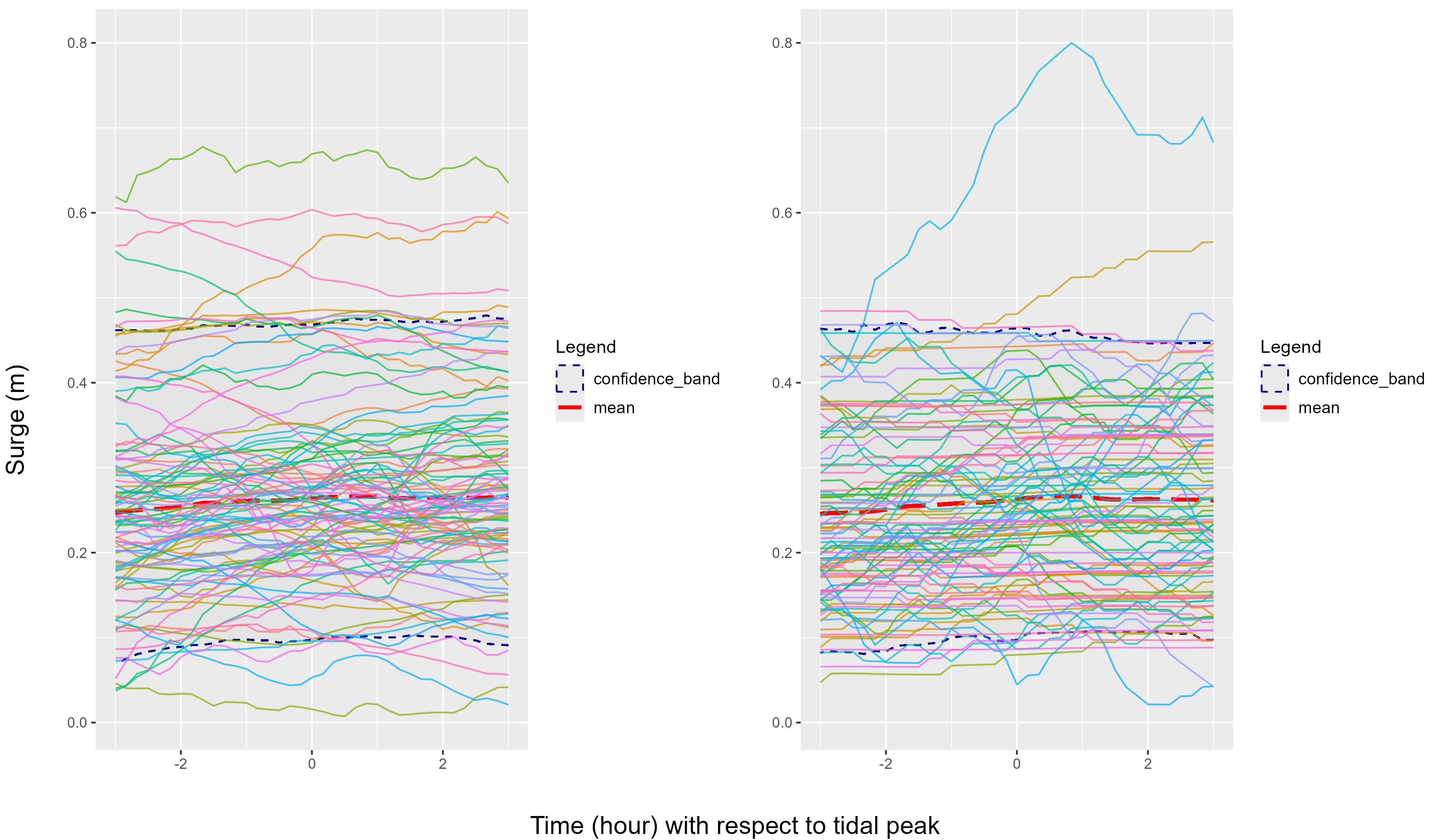}
    \caption{Comparing simulated time series and extreme observations. Left panel: simulated sample of $100$ extreme time series, right panel: sample of $100$ extreme observations. The dotted lines represent the $95\%$ confidence bands}
    \label{fig:simul_vs_real_sample}
\end{figure}
\subsection{Consistency of simulations with extreme observed time series}
We validate our simulator by comparing simulated times series with extreme observations. To this end, we use a set of diagnostic tools, using extreme value analysis, PCA and classification algorithms. 
\subsubsection{Comparison of percentile levels.}
\begin{figure}[H]
    \centering
    \includegraphics[width=0.6\textwidth]{ 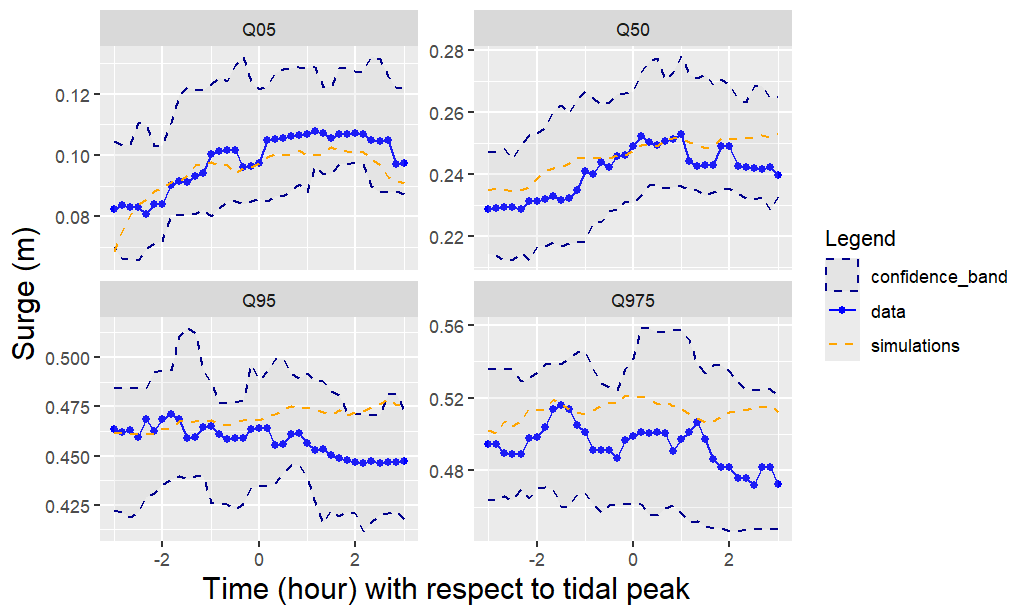}
    \caption{Percentiles obtained in the data and in the simulations (blue lines: data, orange dotted lines: simulations)}
    \label{fig:percent_simul_vs_real_sample}
\end{figure}

First, we compare the empirical percentiles at each time point $t$ with those derived from the simulated time series (Fig.~\ref{fig:percent_simul_vs_real_sample}). To quantify uncertainty in the observations, we construct a bootstrap $95\%$ confidence band by resampling the extreme time series $500$ times. The simulated levels obtained are coherent with the observations as the simulations' percentiles lie within the data confidence band.
\subsubsection{Comparison of coordinates in a PCA basis.}
\begin{figure}[H]
    \centering
    \includegraphics[width=0.55\textwidth]{ 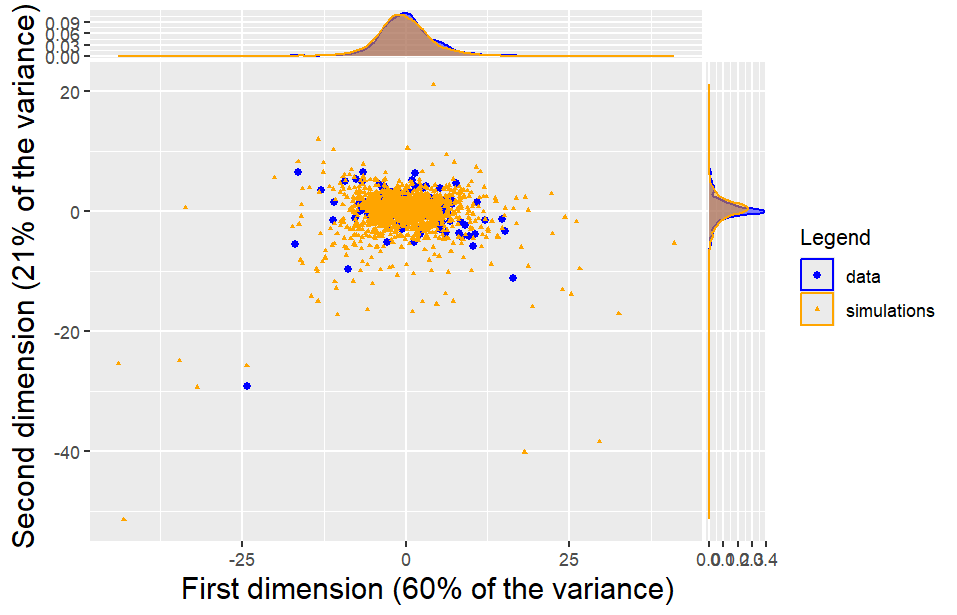}
    \caption{PCA coordinates for S in simulations (orange triangles) and extreme observations (blue dots)}
    \label{sample_coord}
\end{figure}

The simulated extreme time series should exhibit shape similarities with the observed extremes. One simple way to compare the shape of the time series is to use dimensionality reduction methods and to compare the resulting coordinates. To this end, we analyse the angles of the time series as the simulations extrapolate the $L^{2}$ norm of the observations. 
We apply PCA to the normalized series $\tilde{X}/\ell(\tilde{X})$ and, to preserve potential differences, both simulated and observed series are further normalized by using the mean and the standard deviation of the observations. \\
Then, we compare the PCA coordinates of recorded time series with those of simulated time series (Fig.~\ref{sample_coord}). 
The similarity in the score distributions suggests that simulated time series and extreme observations exhibit comparable behavior.
Yet, this similarity is more pronounced in the first dimension, which accounts for the largest proportion of variability (here $60\%$). A KS test confirms that simulations' and observations' coordinates follow the same law in the first dimension ($p$ value$\approx15\%$) but do not in the second dimension ($p$ value $\approx 0.8\%$).
\subsubsection{Comparison of the distribution upper tails.}
\begin{figure}[H]
    \centering
    \includegraphics[width=0.5\textwidth]{ 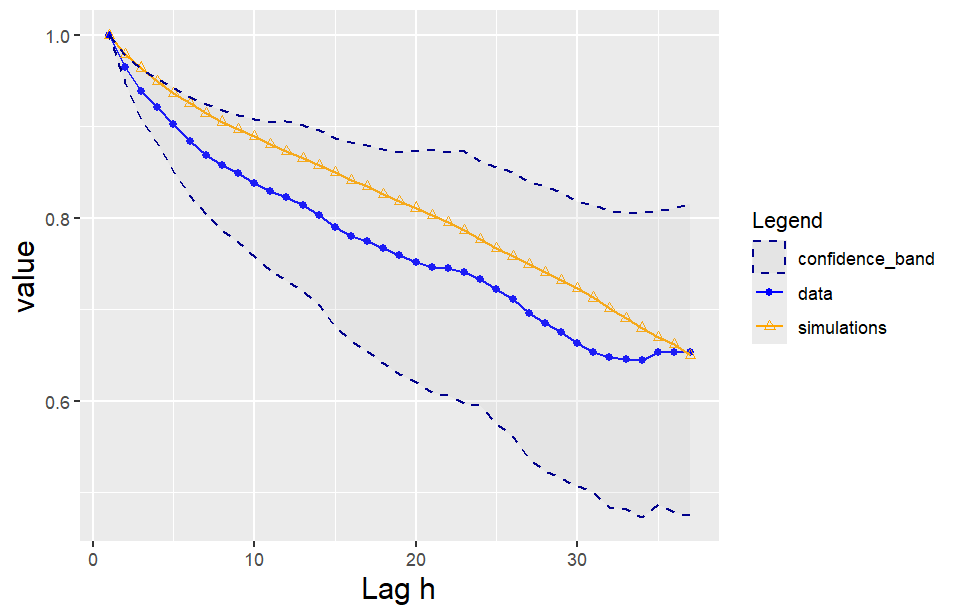}
    \caption{Extremogram for S for simulations (orange) and observations (blue) with confidence bands in dotted lines.}
    \label{fig:extremo_cond}
\end{figure}

Then, we compare the temporal dependence within each tidal cycle between observations and simulations, with a focus on extremal dependence. Following \citep{opitz2021semi, dell2024flexible}, we assume that $\tilde{X}_{M}^t$ and $\tilde{X}_{M}^{s}$ are asymptotically dependent for all pairs $(t,s)$. 
To quantify the dependence, we estimate the $\ell$-extremogram \citep{de2022functional} for both the recorded and simulated time series. We detail the approach in Appendix \ref{corr_extremes}.\\
We focus on the values exceeding the $90\%$ percentile and use $500$ resamples to construct a bootstrap $95\%$ confidence band. The extremogram of the simulated time series lies within the confidence band (Fig.~\ref{fig:extremo_cond}) based on the observed extreme time series. 
Moreover, the density obtained at each time $t$ should be an extrapolation of the observed data. Directly using the quantiles of the simulations may produce very large values. 
This occurs because we simulate $\tilde{X}_{\text{sim}}$ verifying $\ell(\mathcal{T}(\varepsilon_{M}))>u_{\ell}$, meaning their extremes can be larger than those observed. \\
In this setting, we compare our values with the empirical quantiles of the observations and the fitted law of exceedances. 
Following \citet{solari2017peaks}, the results are expressed in terms of return period $P$ and corresponding return levels, with approximately $7$ extreme observations per year (Fig.~\ref{fig:peak}). All implementation details are provided in Appendix~\ref{cplmts_extremes}.\\
The highest simulated levels fall within the confidence band, indicating that our simulations look similar to the extreme observations. Thus, they can be considered a reliable extrapolation of the observed data toward higher values.
\begin{figure}[H]
        \begin{minipage}[c]{0.48\textwidth}
         \includegraphics[width=\textwidth]{ 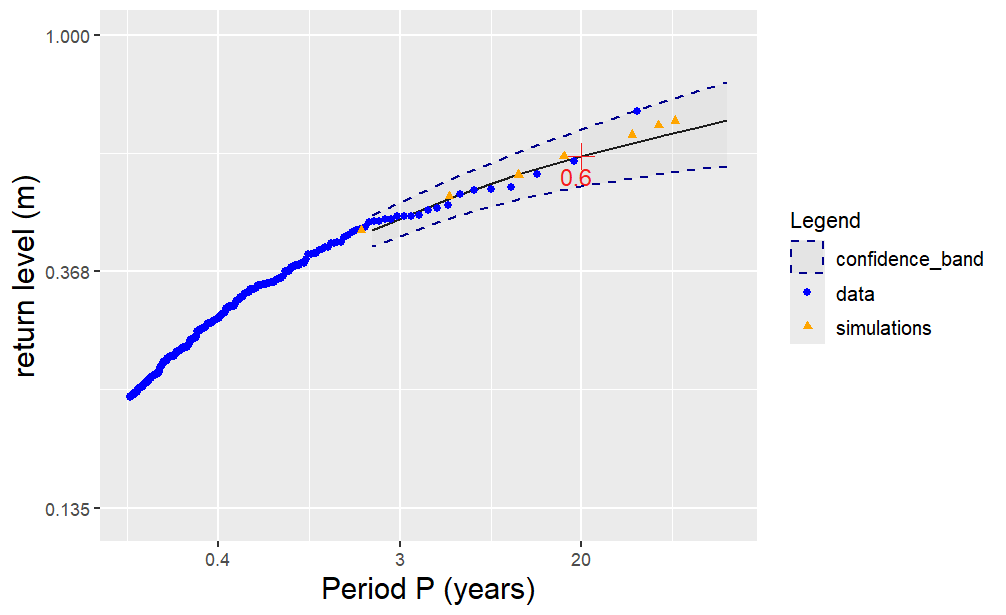}
        \end{minipage}
        \begin{minipage}[c]{0.48\textwidth}
             \includegraphics[width=\textwidth]{ 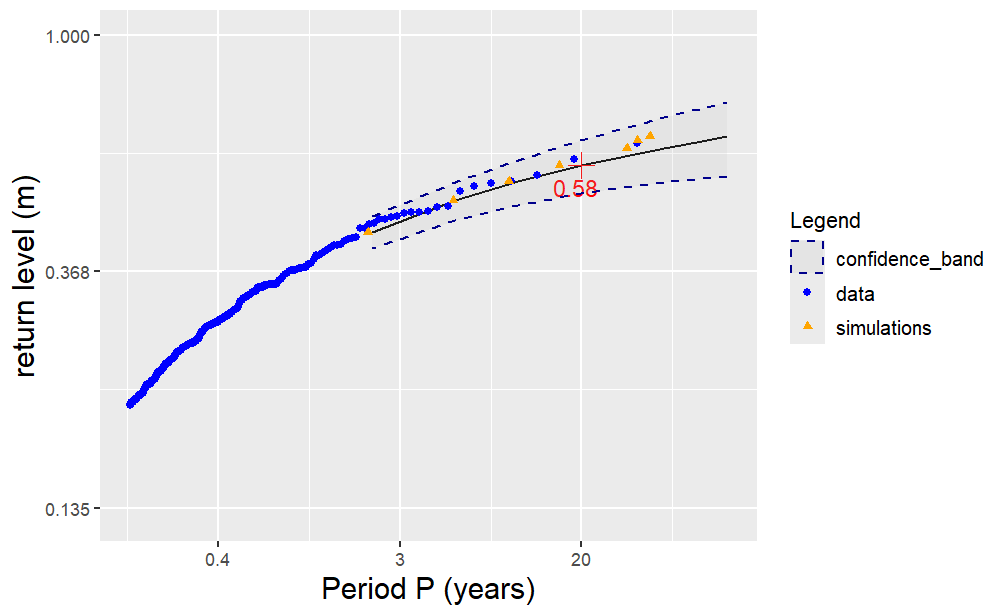}
        \end{minipage}
        \caption{Return level for simulations (orange triangles) and observations (blue dots), for two different values of $t$. Left panel: $t=19$ (tidal peak),  right panel: $t=13$ (one hour before). The dotted lines represent $95\%$ confidence bands.}
        \label{fig:peak}
\end{figure}
\subsubsection{Comparison of predictions with classification algorithms.}
Finally, following the same idea as two-sample classification tests \citep{lopez2016revisiting}, we rely on a classification-based approach to determine whether the generated extreme times series differ from the original dataset. We define a binary classification problem, where
the target variable $y$ equals $1$ if the input $x$ is a simulated time series, $0$ otherwise. The goal is to learn a function $g$ that, given the input $x$, predicts the correct class $y$. If the simulations behave like the extreme observations, such classifier should hardly identify the simulated time series, resulting in an accuracy close to $50\%$.\\ 
To address this question, we begin by randomly drawing $n=259$ simulated time series without replacement (\textit{Step 1}) since the number of simulated time series $n_{sim}=2,000$ differs from $n$. We construct a dataset composed of these sampled time series and extreme observations, ensuring that simulated time series make up $50\%$ of the data. Then, after splitting this dataset into training and testing subsets, we train our model on the training part (\textit{Step 2}). Finally, we evaluate the model's overall accuracy on the test set (\textit{Step 3}).\\
To account for the influence of the sampling used in \textit{Step 1}, we repeat $100$ times these three steps and provide a confidence interval of order $90\%$ for the overall accuracy.
We use three classifiers with this approach: a support-vector machine (SVM) with a radial kernel, a generalized linear model (GLM), and a random forest \citep{hastie2009elements}.\\
We apply this procedure with
different types of input features: the time series, its norm and its angle. 
If we provide the time series, all classifiers achieve an overall accuracy close to $50\%$ (see Table \ref{result_GLM_SVM_Y}), meaning that the ML models struggle to identify simulated time series. This suggests that our simulations are quite consistent with the observed data.\\ 
For example, the distributions of $\ell(\tilde{X}_{\text{sim}})$ and $\ell(\tilde{X}_{M})$ are quite similar (see Appendix~\ref{compar_distrib}), with differences appearing for the largest values as the simulations better explore the extremes of $\ell(S)$.
In contrast, when using the angle as input feature, the models are better able to distinguish between simulations and observations but the accuracy still remains low on the order of 60\% in average.
\begin{table}[!ht]
\centering
\begin{tabular}{lccc}
\hline
Input & SVM & GLM & RForest \\ \hline
        Hyper-params & kernel: radial & link: logit & Nb-trees: 500 \\ \hline
        X & 44-56 & 40-53 & 48-61 \\ \hline
        $\ell$(X) & 43-55 & 43-54 & 43-56 \\ \hline
        A(X) & 49-64 & 40-53 & 59-72 \\ \hline
\end{tabular}
\caption{Confidence intervals at order $90\%$ for the global accuracy}
    \label{result_GLM_SVM_Y}
\end{table}\\
This approach also reveals the influence of $\tilde{X}_{M-\Delta}$ as the attained accuracy depends on the sampling method, which is discussed in Appendix~\ref{uncond_sampling}. The classifiers identify more easily the simulated time series when an unconditional sampling is used. For instance, the overall accuracy exceeds $50\%$ when the time series itself is used as input.

\section*{Conclusions}
Motivated by the modeling of surge-induced coastal flooding, our goal is to develop a simulator for extreme time series. To this end, we align our time series with the framework of regular variations.
As the observations do not meet standard assumptions, we get back to the framework setting by applying an autoregressive model and marginal transformation to the time series. It allows us to obtain a probabilistic model and to develop a simulation method capable of simulating data-like extreme time series, following the same law as extreme observations, but also consecutive extremes.

We apply this method to a hindcast database \cite{origine_donnees} from the site of Gâvres in Brittany, which is considered as a set of observations in our developments. To evaluate the quality of our simulations, we provide several approaches, including a classification task where the goal is to distinguish between observed and simulated time series.
Our results support the validity of our simulation approach as machine learning algorithms struggle to reliably distinguish the two groups. 

This paper focuses on simulating univariate extreme time series. 
Yet, real-world forcing conditions evolve in a multivariate context \citep{rohmer2022partitioning,GOULDBY201415} as other variables such as wind speed also intervene at each time step.
Therefore, extending the approach to the multivariate case \citep{kim2024extremal} represents a promising direction for effectively modeling extreme forcing conditions. 

\section*{Acknowledgement}
Our work has benefitted from the AI Interdisciplinary Institute ANITI. ANITI is funded by the France 2030 program under the Grant agreement n°ANR-23-IACL-0002.
\newpage
\section{Complements on data analysis}\label{Seasonal}
\begin{figure}[H]
    \centering
    \includegraphics[width=0.65\textwidth]{  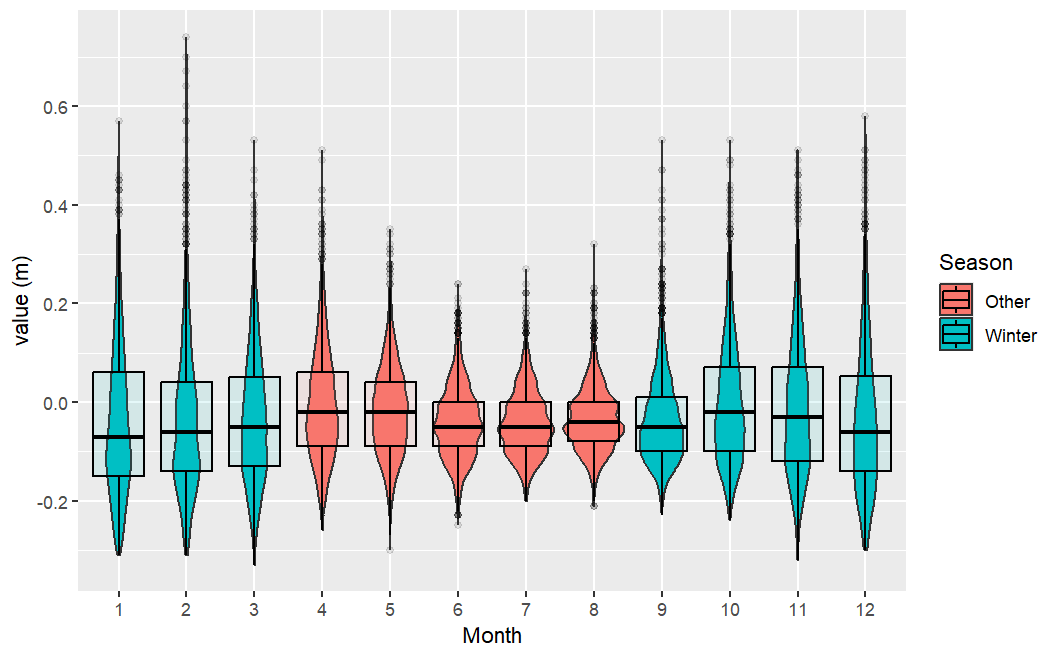}
    \caption{Seasonal evolution of the distribution of $S$ at tidal peak}
    \label{fig:distrib_peak_S_month}
\end{figure}

We present here the evolution of $S$ at tidal peak, $t=19$, according to the month considered (Fig.~\ref{fig:distrib_peak_S_month}). The extremes of winter values are much higher than the extremes of summer values, which is explained by a higher frequency of storms in winter. 
\section{Whitening and marginal transformation}
\subsection{Effect of the autoregressive models}\label{Whitening_}
\begin{figure}[H]
    \centering
    \includegraphics[width=0.6\textwidth]{ 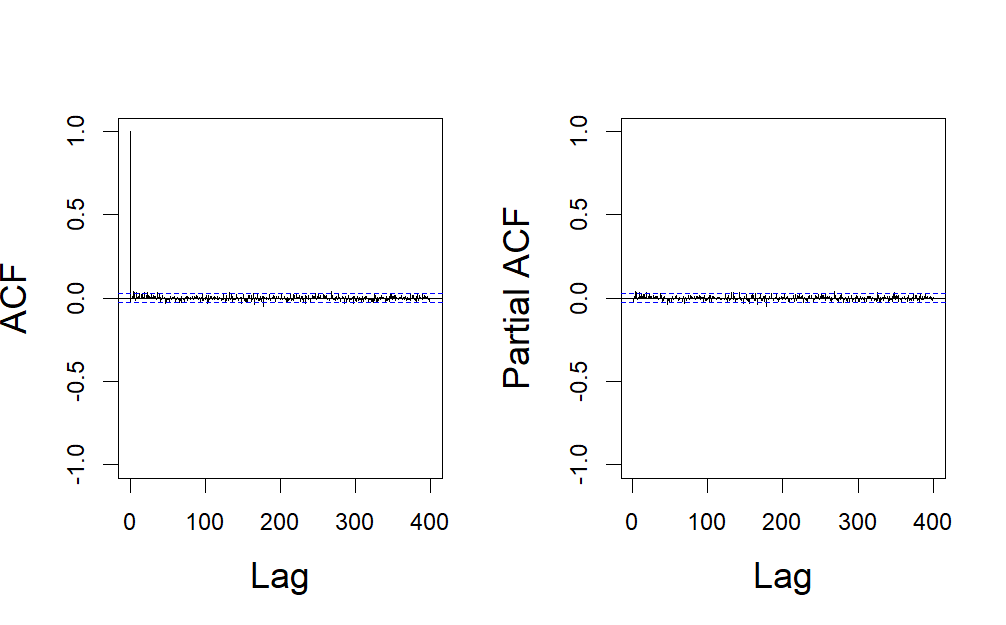}
    \caption{Pearson's correlations observed for S at $t=19$. Left panel: ACF, right panel: PACF}
    \label{fig:acf_pacf_t=19}
\end{figure}

\begin{figure}[H]
    \centering
    \includegraphics[width=0.5\textwidth]{ 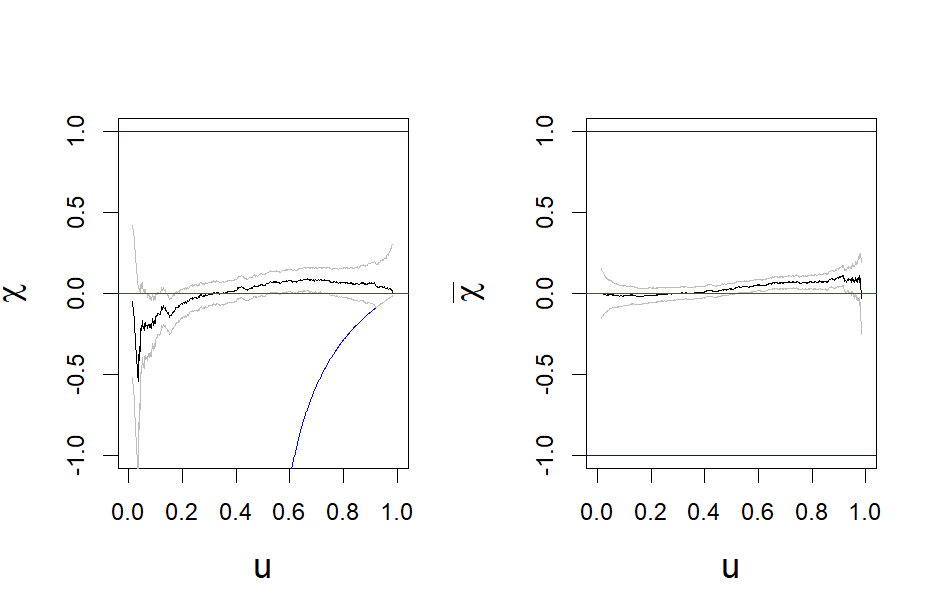}
    \caption{Measures of asymptotic dependence for the surge for ($t=19$, $h=1$) with confidence band (grey lines). The blue lines represent the theoretical bounds of the correlation coefficient.}
    \label{fig:chi_meas}
\end{figure}

We present here the effect of applying the autoregressive models on $\tilde{X}_{M}^{t}$. We can first use the ACF and the PACF on the residuals $\varepsilon_{M}$ (Fig.~\ref{fig:acf_pacf_t=19}). 
\paragraph{Correlations of extremes.}
We recall the definition of the $\chi$  measure \citep{coles2001introduction}. Let $U$ and $V$ be two variables with uniform distributions and we note $\chi(u)=\proba(V>u\mid U>u)$ with $(0<u<1)$. They are asymptotically independent if $\chi(u)\xrightarrow{}0$ when $u\xrightarrow{}1$. In this case, $\Bar{\chi}$ measures the dependence between the two variables.\\
In our case, the pair ($U,V$) corresponds for a given $t$ and $h$ to the pair ($\varepsilon_{M}^{t},\varepsilon_{M+\Delta h}^{t}$) after a rank transformation. At the tidal peak ($t=19$), $0$ belong to the confidence interval of $\chi(u)$ when $u>0.9$ for $h=1$ (Fig.~\ref{fig:chi_meas}). As a consequence, we cannot reject the hypothesis of asymptotic independence.
\subsection{Marginal transformation}\label{GPD_choice}
\begin{figure}[H]
    \centering
    \includegraphics[width=0.7\textwidth]{ 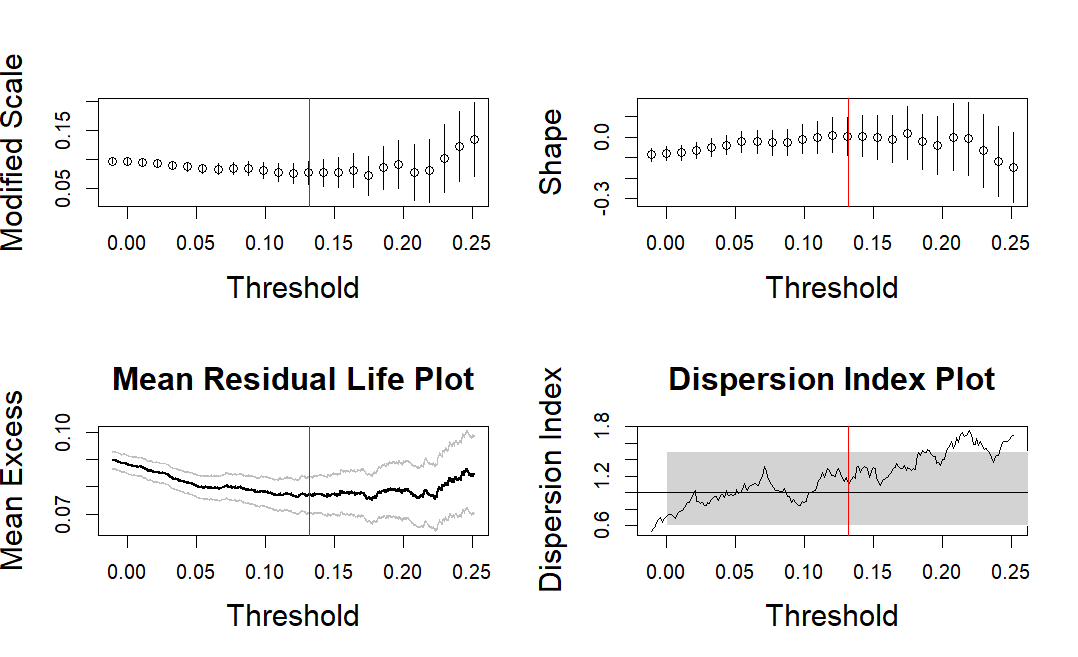}
    \caption{Graphic diagnostics for the residual of $S$ at tidal peak ($t=19$)}
    \label{fig:diag_tidal_peak}
\end{figure}

We describe the evolution of diagnostic statistics when $u^{t}$ is between the median of $\varepsilon_{M}^{t}$ and its $98\%$ quantile. We observe the evolution of the MRL and the updated parameter (Fig.~\ref{fig:diag_tidal_peak}). We represent with the red line the threshold chosen for the marginal transformation. We do have a stable couple ($\sigma',\gamma'$) (top left and top right) and the MRL (lower left corner) seems to be a linear function. 
We show the evolution of the dispersion index in the lower right corner. This tool is based on the link between extreme values and Poisson point process. $u^{t}$ is a consistent threshold if the index is around $1$, which is the case for the threshold we chose. 
\section{Complements on the simulation of extreme time series}\label{cmplts_Cops_Stattest}
\subsection{Choice of the extreme individuals}\label{theta_conv_ul}
\begin{figure}[H]
    \centering
    \includegraphics[width=0.7\textwidth]{ 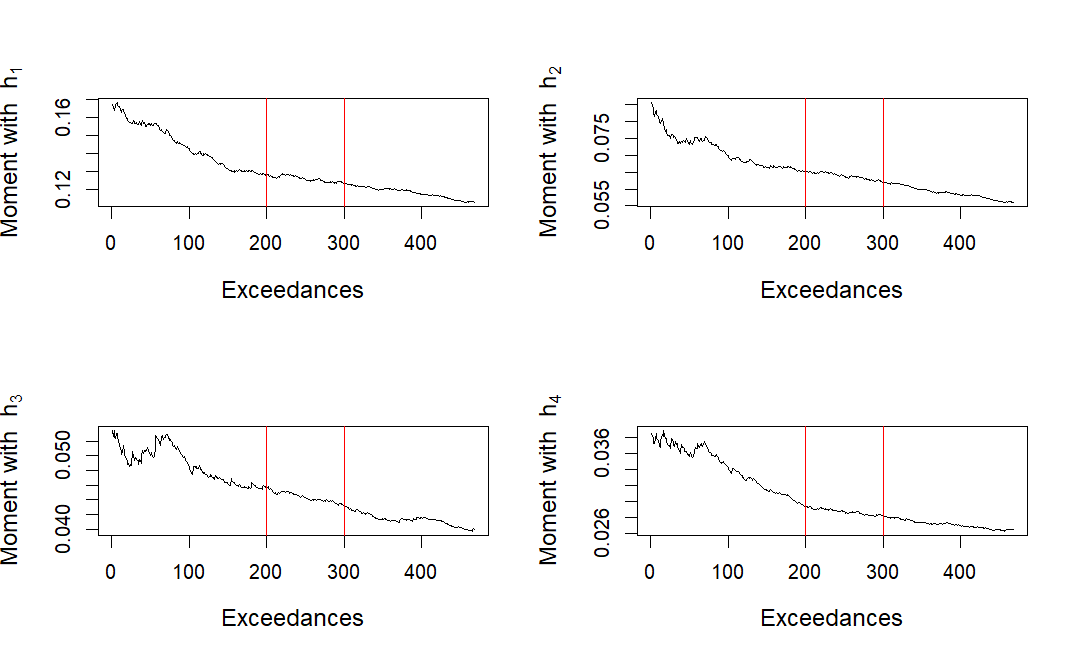}
    \caption{Evolution of the mean projection for $T(S)$}
    \label{convergce_theta}
\end{figure}

The convergence of $\Theta_{M}$ is close in spirit to the stability of $\text{GPD}$ parameters (Sect.\ref{marg_transf_GPD}) as we look for a convergence period of $\mathbb{E}(|\langle \Theta_{M},h_{j}\rangle|)$. 
We use here as $h_{j}(t)=sin(2\pi jt)$ with $j\in \{1,\dots,8\}$. We consider the evolution of the mean of $|\langle f/\ell(f),h_{j}\rangle|\mid \ell(f)>u_{\ell}$ when we increase $u_{\ell}$. Figure ~\ref{convergce_theta} shows that there is a period of stability for the mean projection for several functions when we have between $200$ and $300$ extreme time series. 
\subsection{Choice of the number $J$}
\begin{figure}[H]
    \centering
    \includegraphics[width=0.5\textwidth]{ 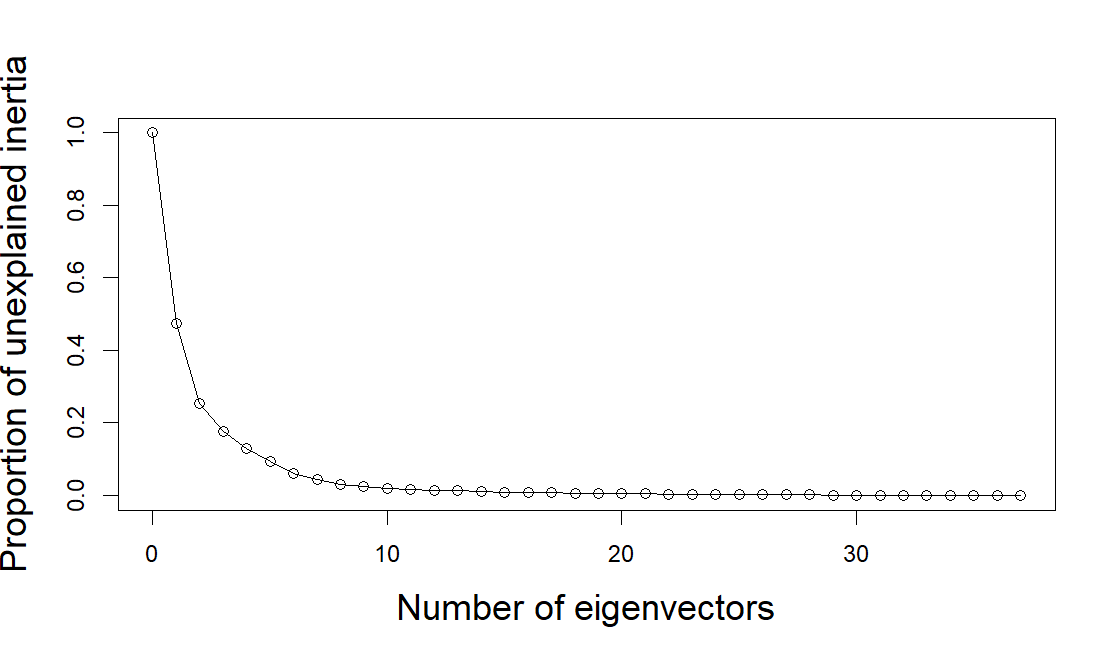}
    \caption{Evolution of the unexplained inertia of $\Theta_{M}$ for $S$}
    \label{fig:prop_inertia}
\end{figure}

\begin{figure}[H]
    \centering
    \includegraphics[width=0.6\textwidth]{ 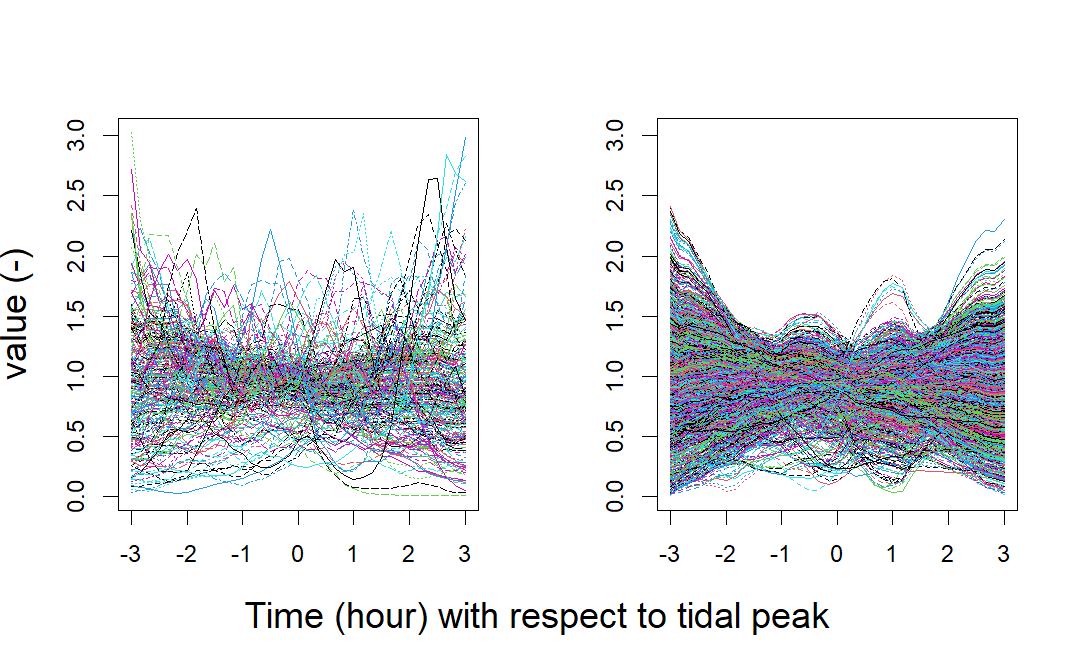}
    \caption{Comparison of the angles $\Theta_{\text{sim}}$ and $\Theta_{M}$ for $S$}
    \label{fig:simul_theta}
\end{figure}

As we increase the number $J$ used in the truncated expression of $\Theta_{M}$, we analyze the evolution $1-R_{J}$ (Fig.~\ref{fig:prop_inertia}). 
After choosing $J=3$, we simulate $n_{sim}=2,000$ angles and compare our simulations with the angles of the extreme observations (Fig.~\ref{fig:simul_theta}). 
\subsection{Details on the copula family}\label{details_cop}
\begin{figure}[H]
    \begin{minipage}[c]{0.48\textwidth}
    \includegraphics[width=\textwidth]{ 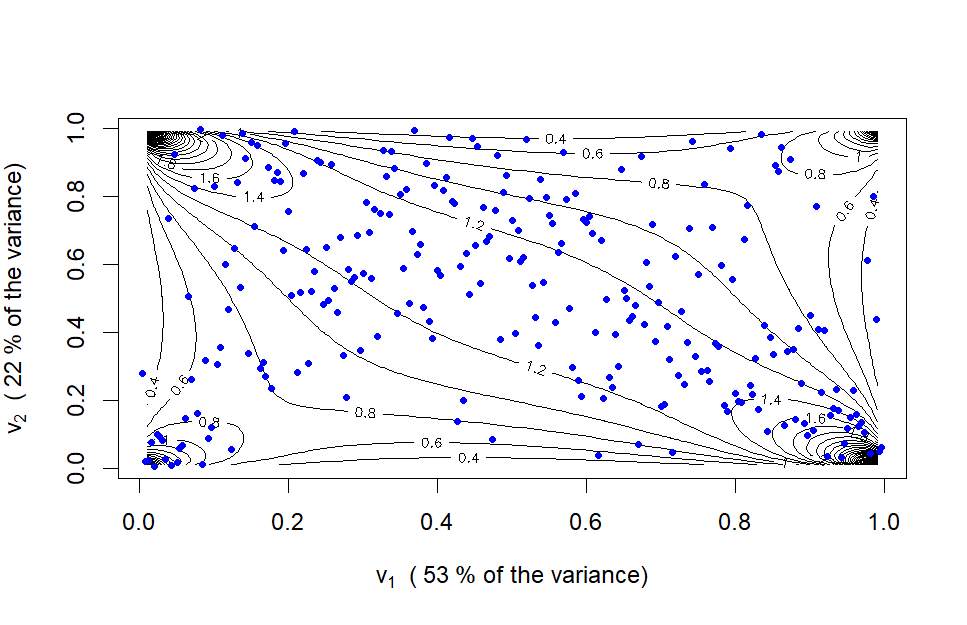}
    \end{minipage}
    \begin{minipage}[c]{0.48\textwidth}
    \includegraphics[width=\textwidth]{ 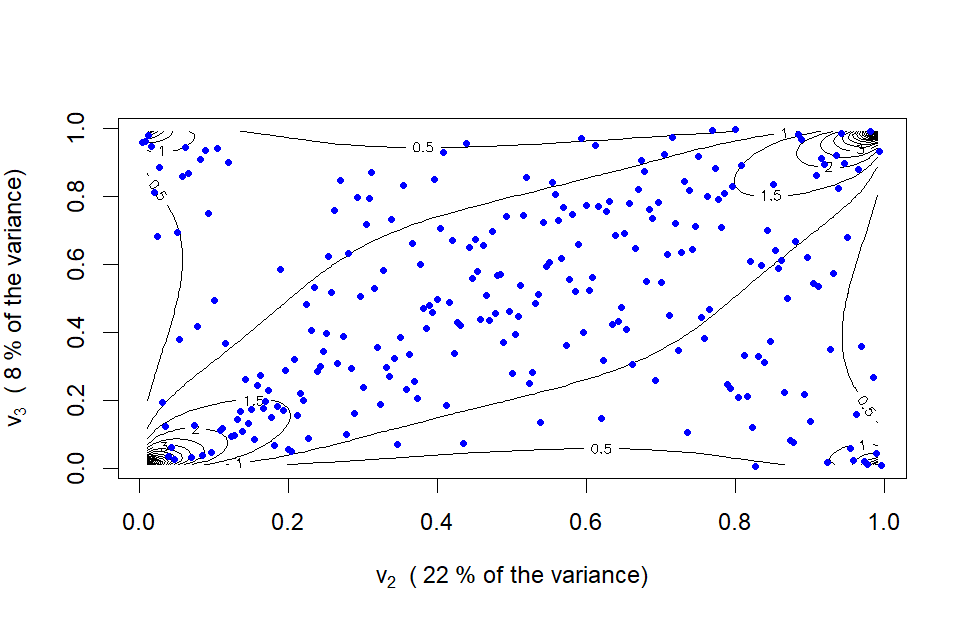}
    \end{minipage}
    \caption{Iso-density curves (dark lines) of the copula with the data coordinates (blue points): (a) $(v_{1},v_{2})$, (b) $(v_{2},v_{3})$}
    \label{model_vs_data_cop}
\end{figure}

\begin{figure}[H]
    \begin{minipage}[c]{0.48\textwidth}
    \includegraphics[width=\textwidth]{ 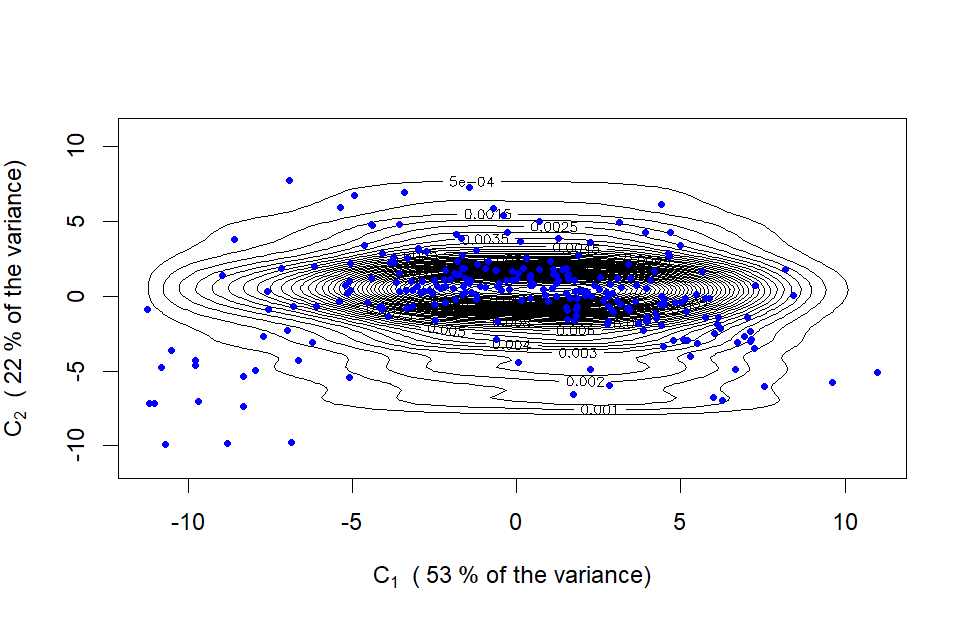}
    \end{minipage}
    \begin{minipage}[c]{0.48\textwidth}
    \includegraphics[width=\textwidth]{ 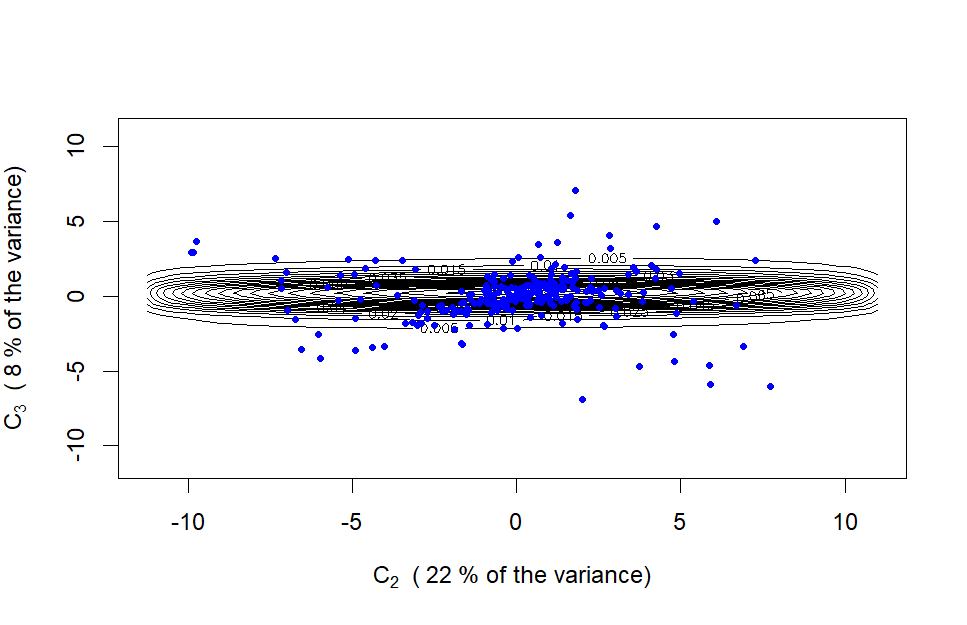}
    \end{minipage}
    \caption{Iso-density curves (dark lines) of the joint law with the data coordinates (blue points): (a) $(C_{1},C_{2})$, (b) $(C_{2},C_{3})$}
    \label{model_vs_data}
\end{figure}

We describe here the parameters estimated for the copula (Fig.~\ref{fig:copules_trees}). 
\begin{table}[H]
    \centering
    \begin{tabular}{lccccccc}
    \hline
        \textbf{Tree} & \textbf{Couples-cond} & \textbf{Name} & \textbf{par} & \textbf{par2} & $\tau$ & $\lambda_{+}$ & $\lambda_{-}$ \\ \hline
        1 & 1,2 & t & -0.2 & 2 & -0.2 & 0.1 & 0.1 \\ \hline
        1 & 2,3 & t & 0.3 & 2 & 0.2 & 0.3 & 0.3 \\ \hline
        2 & 1,3;2 & t & -0.2 & 3.4 & -0.1 & 0.1 & 0.1 \\ \hline
    \end{tabular}
    \caption{Modeling of $A(f)$'s coordinates with copulas: the first row explains the modeling of the law of the first two coordinates while the last one presents the modeling of the conditional law of $(C_{1},C_{3})$ knowing $C_{2}$}
    \label{fig:copules_trees}
\end{table}
We compare the fitted copula models with observed coordinates by looking at the relationship between the two dimensions. In Fig.~\ref{model_vs_data_cop} and Fig.~\ref{model_vs_data}, we compare the data observations with the iso-density curves of the model respectively in the uniform scale and in the data scale. \\Figure~\ref{model_vs_data_cop} shows that we are able to capture the most of the dependence structure as the repartition of data points is coherent with the levels found. However, if we return to the original scale of the coordinates, we see on Fig.~\ref{model_vs_data} the limits of our modeling as some data points do not coincide with the iso-density curves of our model.  
We also use statistical test to compare our model with the law of the observations. The null hypothesis $\text{H}_{0}$ is that the observations follow the model. We reject $\text{H}_{0}$ if we apply the White test under the asymptotic distribution of the test statistic ($p$ \text{value}$\approx 0.2\%$). However, when using the bootstrapped distribution, we do not reject the null hypothesis, which is consistent with the result of the Kolmogorov-Smirnov (KS) test on the empirical copula process. 
\subsection{Choice of $X_{M-\Delta}$}\label{choice_previous_obs}
\begin{figure}[H]
    \centering
    \includegraphics[width=0.5\textwidth]{ 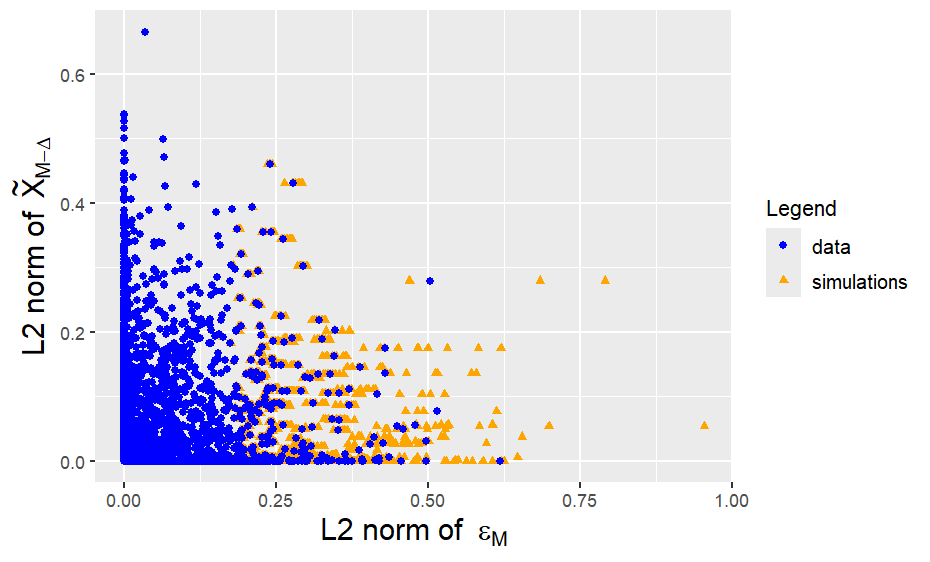}
    \caption{Relation between $\ell(\varepsilon_{M})$ and $\ell(\tilde{X}_{M-\Delta})$ for $S$ (blue dots: data points, orange triangles: coordinates obtained in the simulations)}
    \label{fig:choice_yo}
\end{figure}

We apply a conditional sampling of $X_{M-\Delta}$ knowing $\varepsilon_{sim}$. The resulting levels are quite coherent with those observed in the dataset (Fig.~\ref{fig:choice_yo}) and simulations provide as expected fill gaps where data observations are not available.
\section{Complements on the consistency of simulations with the observations}
\subsection{Correlation of extremes}\label{corr_extremes}
\begin{figure}[H]
    \centering
    \includegraphics[width=0.6\textwidth]{  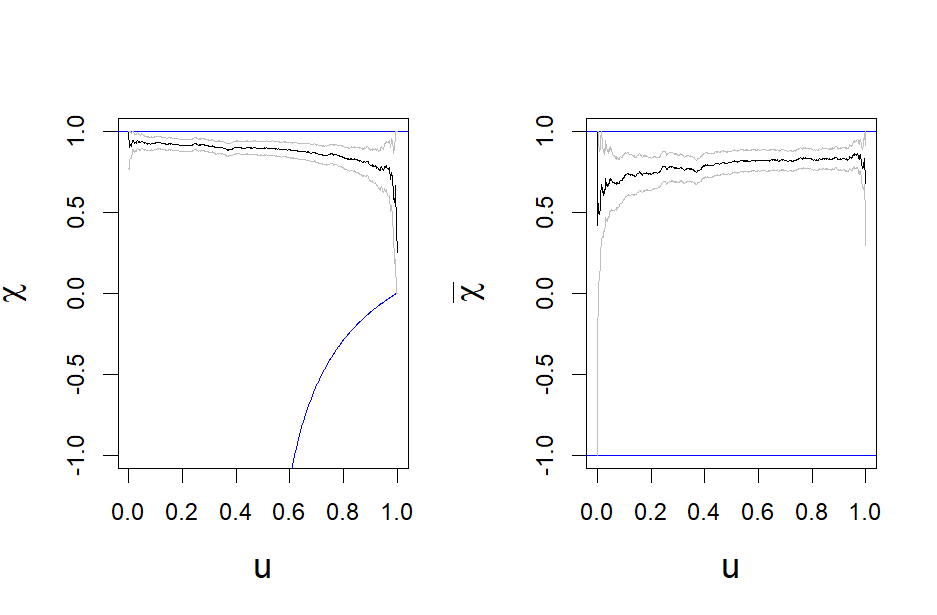}
    \caption{Measures of asymptotic dependence between $\tilde{X}_{M}^{1}$ and $\tilde{X}_{M}^{37}$ for the surge with confidence band (grey lines). The blue lines represent the theoretical bounds of the correlation coefficient.}
    \label{fig:asymp_depend_chi}
\end{figure}

The extremogram is defined for extreme time series with 
\begin{align}
    \label{eq_extremo}
    \pi(s',s)=lim_{q\xrightarrow{}1}\proba(\tilde{X}_{M}^{s'}>u_{q}(s')\mid \tilde{X}_{M}^{s}>u_{q}(s))
\end{align} 
where $u_{q}(s)=F_{s}^{-1}(q)$ is the quantile function of order $q$ of $\tilde{X}_{M}^{s}$. We use $q=0.9$ to compare our simulations with the extreme time series. This tool is based on the hypothesis of asymptotic dependence between the values of the same tidal cycle. We can see for example that the first and the last value of a time series are asymptotically dependent (Fig.~\ref{fig:asymp_depend_chi}) as $\chi$ does not converge to $0$ when $u$ goes to $1$. 
\subsection{Return level estimation}\label{cplmts_extremes}
To obtain the levels obtained in the simulations, we use the formula  
\begin{align}
    \label{eq_cond_double}
   \proba(A\mid B)=\proba(A\cap C\mid B)+\proba(A\cap \Bar{C}\mid B) 
\end{align}
with $A=(\tilde{X}_{M}^{t}\leq x)$, $B=(\tilde{X}_{M}^{t}>u^{t})$ and $C=(\ell(\mathcal{T}(\varepsilon_{M}))>u_{\ell})$. The first member on the right-hand side is directly estimated in the simulated time series whereas we use observations to estimate the second member. Then, we use the model quantiles as $x$ levels. The probability $p=\proba(A\mid B)$ is associated to a return period. 
The link between the quantile order $p$ and the return period $P$ is given by the formula 
\begin{align}
    \label{meaning_return_period}
    P = \frac{1}{\text{npy}(1-p)}
\end{align} 
where $\text{npy}$ is the average number of extreme observations per year, chosen at $7$ events per year. Thus, looking at Fig.~\ref{fig:peak} (left panel), if we observe one day a surge of $0.6\text{m}$  at tidal peak ($t=19$ for every cycle), we will wait in average $20$ years before seeing an event reaching this extreme level.
\subsection{Extreme aspects}
\begin{figure}[H]
        \begin{minipage}[c]{0.48\textwidth}
            \includegraphics[width=\textwidth]{ 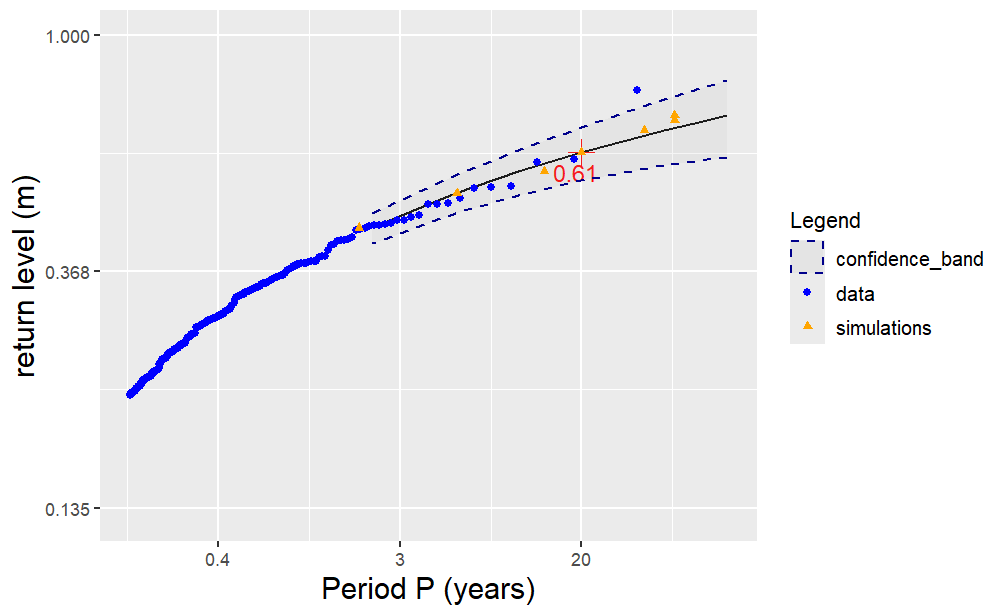}
        \end{minipage}
        \begin{minipage}[c]{0.48\textwidth}
             \includegraphics[width=\textwidth]{ 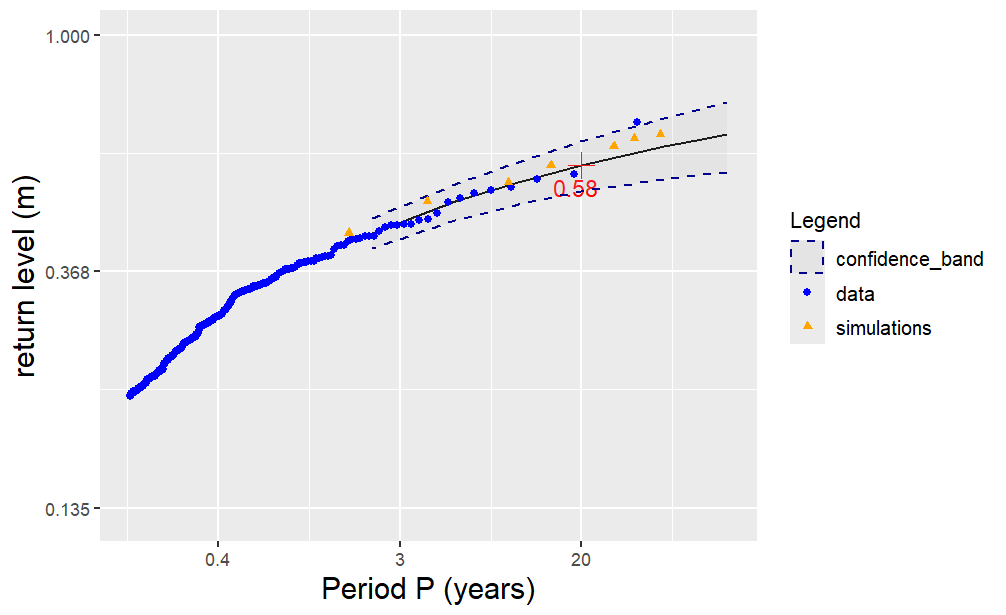}
        \end{minipage}
        \caption{Return level for simulations (orange triangles) and observations (blue dots), for two different values of $t$. Left panel: $t=25$ (one hour after the tidal peak),  right panel: $t=31$ (two hours after the peak). The dotted lines represent $95\%$ confidence bands.}
        \label{fig:peak_2}
\end{figure}

We can analyze the extreme values obtained for other values of $t$ in the simulations and in the data (Fig.~\ref{fig:peak_2}). The levels obtained are consistent with the theoretical values. 

\subsection{Distribution of $\ell(.)$ for simulated and observed extreme time series}\label{compar_distrib}
\begin{figure}[H]
    \centering
    \includegraphics[width=0.6\textwidth]{ 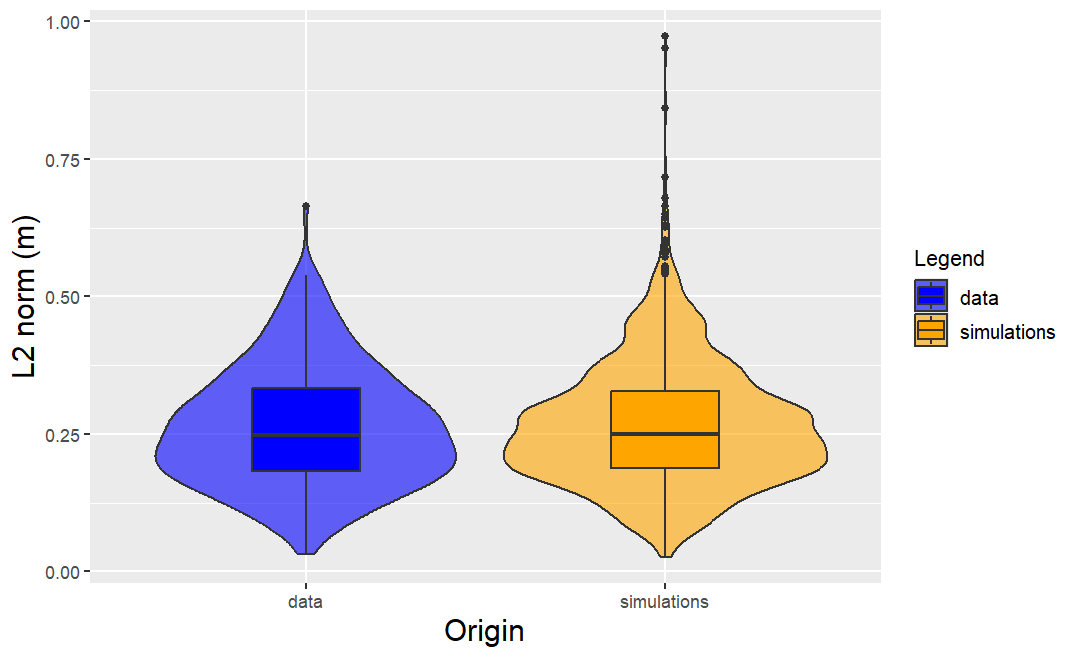}
    \caption{Distribution of $\ell(S)$ (orange: simulated time series, blue: recorded extreme time series)}
    \label{L2_simul_obs}
\end{figure}

We describe here what we obtain for the $\ell(.)$ function in the simulations and in the extreme time series (Fig.~\ref{L2_simul_obs}). The shape of the two distributions are quite similar but their extreme values are quite different. While the maximum value in the data is slightly above $0.6$m, the maximum value obtained in the simulation is larger than $0.8$m. 

\section{Choice of $\tilde{X}_{M-\Delta}$}\label{uncond_sampling}
Here, we present the effect of using an unconditional sampling of $\tilde{X}_{M-\Delta}$.
\subsection{Comparison of percentile levels}
\begin{figure}[H]
    \centering
    \includegraphics[width=0.7\textwidth]{   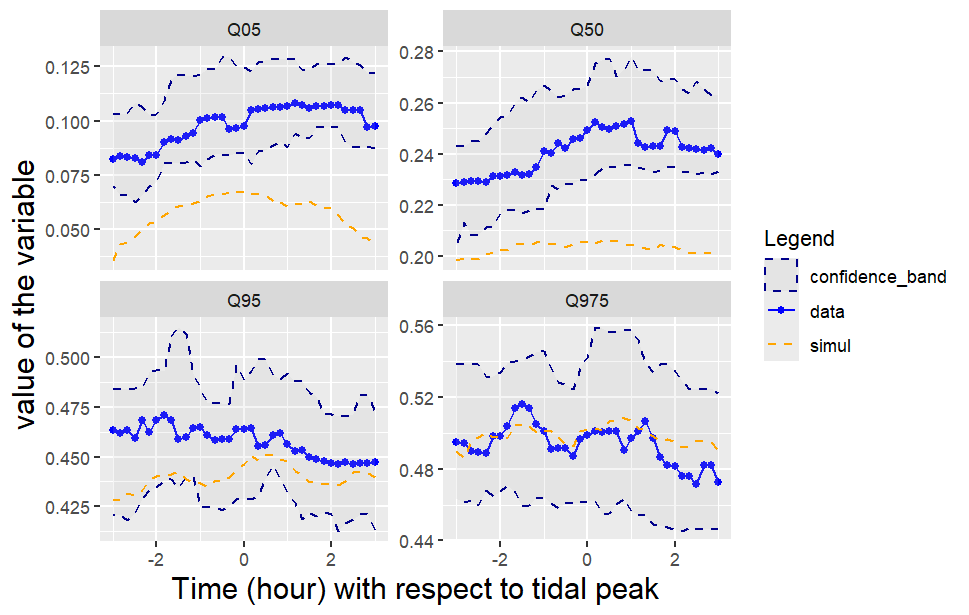}
    \caption{Percentiles obtained in the data and in the simulations without the conditional sampling (blue lines: data, orange dotted lines: simulations)}
    \label{fig:choice_S_FREE}
\end{figure}

Firstly, we analyze the quantiles obtained for each value of $t$ in the simulated time series and in the extreme observations (Fig.~\ref{fig:choice_S_FREE}). We see that for several orders the quantile obtained in the simulations do not belong to the confidence interval of the extreme observations. 
\subsection{Comparison of coordinates in a PCA basis}
\begin{figure}[H]
    \centering
    \includegraphics[width=0.6\textwidth]{   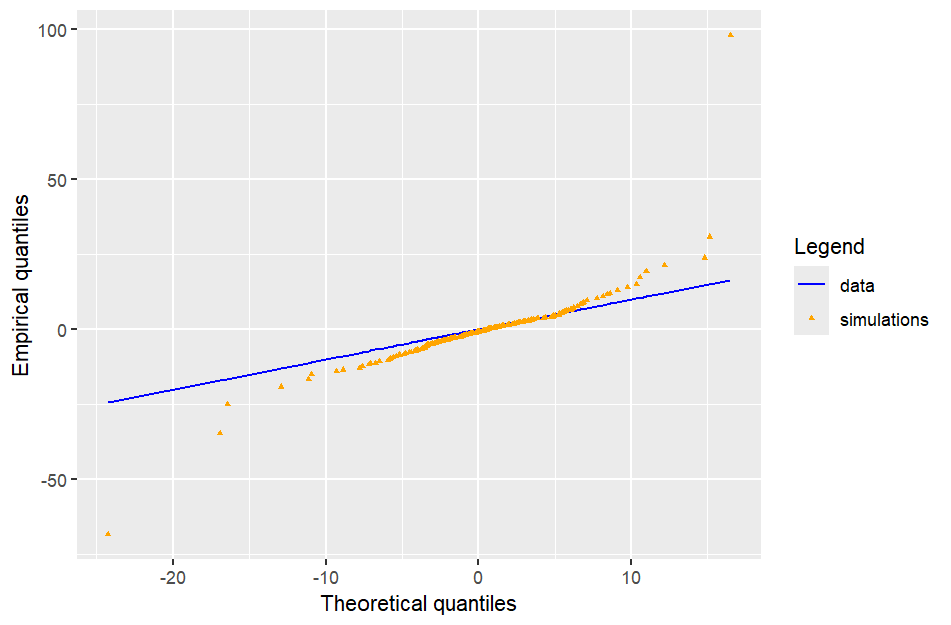}
    \caption{Q-Q plot for $C_{sim,1}$ (orange triangles) on data scale (blue line) with the unconditional sampling of $\Tilde{X}_{M-\Delta}$}
    \label{fig:sample_coord_FREE_1}
\end{figure}

We use the PCA decomposition of the angle $\Tilde{X}/\ell(\Tilde{X})$ of the extreme observations with the unconditional sampling of $\Tilde{X}_{M-\Delta}$. Figure \ref{fig:sample_coord_FREE_1} shows that the quantiles of simulated coordinates do not behave like the observed coordinates as empirical quantiles are deviating from the bisector. As a consequence, the observations' and simulations' coordinates do not follow the same law for the first dimension with a $p$ value of $0.01\%$ for the KS test. 

\subsection{Comparison of the distribution upper tails}
\begin{figure}[H]
    \centering
    \includegraphics[width=0.5\textwidth]{   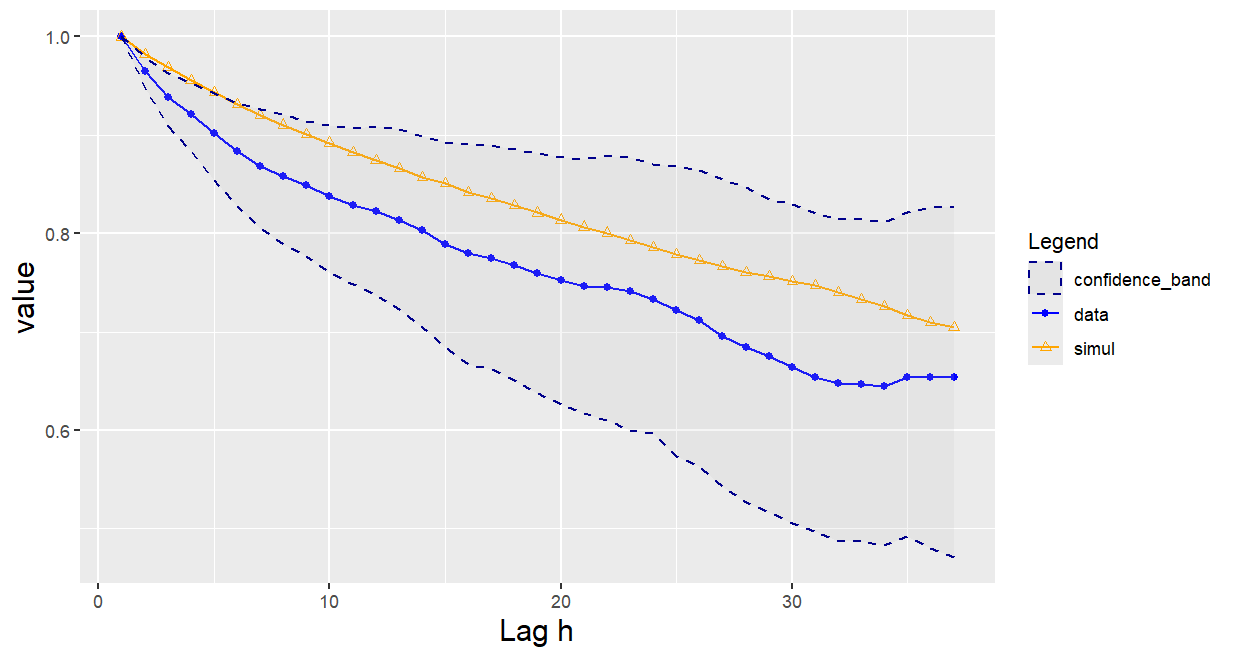}
    \caption{Extremogram for S for simulations (orange) and observations (blue) with confidence bands in dotted lines. Values obtained with the unconditional sampling}
    \label{extremogram_X_S_FREE}
\end{figure}

We compare the behaviour of extreme values for the simulated and recorded extreme time series when unconditional sampling is used. We see for instance that the $\ell$-extremogram estimated in the simulations is further from the data estimation with conditional sampling than what we obtain with conditional sampling (Fig.~\ref{extremogram_X_S_FREE}). \\
Consequently, the selection method does have an impact on the result. 

\subsection{Comparison of predictions with classification algorithms}
We focus on the performances of the classifiers (Table \ref{result_GLM_SVM_Y_uncond}). The models are able to identify the simulated time series as $50\%$ does not belong to the confidence interval.  
\begin{table}[!ht]
    \centering
    \begin{tabular}{lccc}
    \hline
        Input & SVM & GLM & RForest \\ \hline
        Hyper-params & kernel : radial & link : logit & Nb-trees : 500 \\ \hline
        X & 53-65 & 49-62 & 51-65 \\ \hline
        l(X) & 51-62 & 50-62 & 46-56 \\ \hline
        A(X) & 54-65 & 47-58 & 62-72 \\ \hline
    \end{tabular}
    \caption{Confidence intervals at order $90\%$ for the global accuracy}
    \label{result_GLM_SVM_Y_uncond}
\end{table}
  
\bibliographystyle{plainnat}
\bibliography{ma_biblio.bib}
\end{document}